\documentclass[
  aps,
  prx,              
  reprint,          
  superscriptaddress,
  longbibliography,  
  floatfix
]{revtex4-2}

\usepackage{graphicx}
\usepackage{dcolumn}
\usepackage{bm}
\usepackage[colorlinks=true, allcolors=blue]{hyperref}
\usepackage{physics}
\usepackage{subfig}

\usepackage{subcaption}
\usepackage{booktabs}
\usepackage{amssymb}
\usepackage{caption}
\usepackage{svg}

\usepackage{comment}

\begin{document}

\preprint{APS/123-QED}

\title{Quantum physics informed neural networks for multi-variable partial differential equations}

\newcommand{\Unimi}{%
  Dipartimento di Fisica “Aldo Pontremoli”, Università degli studi di Milano,\\
  Via G. Celoria, 16, 20139 Milano, Italy%
}

\author{Giorgio Panichi}
\affiliation{\Unimi}
\affiliation{%
  Istituto di Fotonica e Nanotecnologie, Consiglio Nazionale delle Ricerche,\\
  Piazza Leonardo da Vinci 32, 20133 Milano%
}

\author{Sebastiano Corli}
\affiliation{\Unimi}

\author{Enrico Prati}
\email{enrico.prati@unimi.it}
\affiliation{\Unimi}
\affiliation{%
  Istituto di Fotonica e Nanotecnologie, Consiglio Nazionale delle Ricerche,\\
  Piazza Leonardo da Vinci 32, 20133 Milano%
}

\date{\today}

\begin{abstract}
Quantum Physics-Informed Neural Networks (QPINNs) integrate quantum computing and machine learning to impose physical biases on the output of a quantum neural network, aiming to either solve or discover differential equations. The approach has recently been implemented on both the gate model and continuous variable quantum computing architecture, where it has been demonstrated capable of solving ordinary differential equations. Here, we aim to extend the method to effectively address a wider range of equations, such as the Poisson equation and the heat equation.

To achieve this goal, we introduce an architecture specifically designed to compute second-order (and higher-order) derivatives without relying on nested automatic differentiation methods. This approach mitigates the unwanted side effects associated with nested gradients in simulations, paving the way for more efficient and accurate implementations.

By leveraging such an approach, the quantum circuit addresses partial differential equations -- a challenge not yet tackled using this approach on continuous-variable quantum computers. As a proof-of-concept, we solve a one-dimensional instance of the heat equation, demonstrating its effectiveness in handling PDEs, both in an ideal and a noisy regime. We report our experiment on a photonic hardware to address a realistic noise scenario for our simulations. Such a framework paves the way for further developments in continuous-variable quantum computing and underscores its potential contributions to advancing quantum machine learning.

\end{abstract}

\maketitle


\section{\label{sec:intro}Introduction}

Quantum Physics-Informed Neural Networks (QPINNs) integrate quantum computing with classical Physics-Informed Neural Networks (PINNs), combining the strengths of quantum computation~\cite{feynman2018simulating,lloyd1999quantum,raissi2019physics,Killoran2019,knudsen2020solving} and machine learning~\cite{prati2017quantum,biamonte2017quantum,corli2024quantum}. By merging these fields, QPINNs have the potential to surpass classical computational limitations, particularly in the addressing of complex physics and engineering problems~\cite{schuld2015introduction,raissi2019physics}. Realizing QPINNs requires the construction of quantum neural networks, which can be implemented through various quantum computing architectures, including qubit-based and continuous-variable quantum computing (CVQC).
Using the fundamental building blocks of CVQC, Continuous Variable Quantum Neural Networks (CVQNN) emulate classical neural networks by allowing circuit parameters, the gate parameters, to be optimized using classical algorithms~\cite{Killoran2019}.

CVQC relies on the quadratures of the electromagnetic field, whose measurement provides a way to perform analog computing on a quantum device. The framework of photonic computing offers many advantages, for instance photons are not subject to decoherence, and computation may be run at room temperature~\cite{baldazzi2025four}. These assets help photonic computing to be portable in harsh environments, such as satellites~\cite{torta2025quantum,yin2017satellite,goswami2023satellite,ren2017ground} or underwater vehicles~\cite{yan2021underwater,sun2020review}.

Complementing CVQC, physics-informed neural networks (PINNs), initially introduced by Raissi et al.~\cite{raissi2019physics}, incorporate established physical laws directly into the training loss function. These physical laws, typically formulated as differential equations, guide the neural network towards data-consistent solutions that inherently respect governing physical principles~\cite{Raissi2019,brevi2024tutorial}. PINNs have demonstrated effectiveness in solving complex partial differential equations (PDEs), especially in scenarios such as high-dimensional problems~\cite{Hu_2024}, irregular geometries~\cite{xiang2022hybridfinitedifferencephysicsinformed}, non-integrable systems~\cite{brevi2024addressing}, or situations in which finite-difference and mesh methods become impractical~\cite{jin2021nsfnets,hu2023applying,gu2024physics}. Further, notable approaches like the unified PDE-solving strategy by Berg and Nyström~\cite{berg2018unified}, and the DeepXDE library by Lu et al.~\cite{lu2021deepxdex}, reinforce the potential of PINNs.

A paramount key feature of PINNs relies in their integration of empirical data with the solution of the differential equation. Leveraging such additional information, neural networks have proven capable to infer the unknown parameters of the differential equations -- e.g. the pulsation $\omega$ for the harmonic oscillator. When transposing classical PINNs on a quantum hardware, quantum data could potentially drive to the solution of Schr{\"o}dinger equations. In the future, quantum sensing techniques could be integrated to resolve the wave function of many-bodies systems, which are not classically tractable.

Combining CVQC and PINNs naturally leads to QPINNs, which exploit quantum computational advantages for physics-informed machine-learning tasks. First attempts to exploit this potential include applications such as solving one-dimensional Poisson equations using single-qumode circuits~\cite{Markidis2022} and employing multi-qumode structures for deeper networks in various differential equations~\cite{knudsen2020solving}. Furthermore, PINNs have been implemented on qubit-based quantum architectures, solving a diverse array of differential equations~\cite{e26080649,sedykh2024hybrid,Siegl2024}, including fully quantum~\cite{leong2022variational,gaitan2020finding,paine2023physics} and hybrid approaches~\cite{trahan2024quantum}.

Despite these advances, existing QPINN implementations typically rely on nested gradient calculations to compute higher-order derivatives. This dependence increases computational complexity and can compromise solution accuracy, restricting the effectiveness of QPINNs in solving more complex differential equations.
Previously, some of the authors have already addressed quantum machine learning tasks by exploiting gate model architecture \cite{lazzarin2022multi}, adiabatic architecture \cite{moro2023anomaly,noe2024quantum} and measurement-based architecture \cite{corli2023max}. Here, to address the above limitation of QPINNs over the continuous variable encoding, we propose a novel quantum neural network architecture specifically designed for computing higher-order derivatives , avoiding nested gradient calculations. Our implementation leverages established quantum computing and machine learning libraries, notably TensorFlow and Strawberry Fields~\cite{tensorflow2015-whitepaper,strawberryfields2024quantum,killoran2019strawberry}, enabling more efficient and accurate solutions. We demonstrate the effectiveness of our proposed QPINN architecture by solving the temporal evolution of the heat equation as a representative PDE.

A rigorous analysis on the quantum speedup, when dealing with finite difference method, has already been carried out by Linden et al.~\cite{linden2022quantum}. The authors, benchmarking five classical methods versus five quantum methods to solve the heat equation in a rectangular domain, state that there is an almost quadratic speedup, for a number of dimension $d\geq 2$. However, in this paper we do not address finite difference methods, but instead a heuristic one, based on variational quantum circuits.

Our approach mitigates current limitations in existing simulation libraries and represents a significant step toward solving broader classes of PDEs with higher precision~\cite{virtanen2020scipy}.

In order to address the robustness against noise, we also develop a realistic noise environment based on an experimental campaign of photon counting carried out on a quantum photonic processor provided by Xanadu company, known as X8 \cite{xanaduX8}.
We exploit such measurements to compute the attenuation factor on each channel and simulate QPINNs in a noisy environment.
The results prove the resilience of QPINNs against optical noise.

This work is organized as follows: Section 2 summarizes the encoding of neural networks over the continuous variable scheme, Section 3 involves our design of the quantum physics informed neural network, while Section 4 discusses our results, Section 5 evaluates resilience of QPINNs with respect to realistic noise.
Finally, in the Conclusion Section we outline implications and future directions.

\section{\label{sec:architecture}Encoding a neural network on a continuous variable quantum computing architecture}

In this Section, we provide the framework for building a neural network, by using the set of quantum gates natively provided by the CV architecture.
More specifically, we provide the methods required to build QPINN including encoding, output measurements and simulation constraints.
Continuous-variable quantum computing (CVQC) encodes information in the continuous degrees of freedom of the electromagnetic field. Specifically, computations are performed by using the canonical quadratures of the electric field, $\hat X$ and $\hat P$, which obey Heisenberg's uncertainty principle, meaning that they cannot be measured simultaneously with arbitrary precision~\cite{lloyd1999quantum,hanamura2024implementing}. In a phase-space representation, the spectra of $\hat X$ and $\hat P$ -- the pair $(x,p)$ -- form a fundamental unit of quantum information called qumode. The use of continuous variables enables floating-point computations, realized through a universal gate set comprising Gaussian operations (such as displacement, squeezing, Fourier transforms, and beam splitters), cubic phase gates, and measurement-induced gates~\cite{takeda2017universal,hillmann2020universal,bangar2023experimentally}. Although Gaussian gates can be efficiently simulated on classical hardware~\cite{ghose2007non}, analogously to Clifford gates in qubit-based quantum computing~\cite{anders2006fast,gottesman1998heisenberg,fujii2015quantum,nielsen2001quantum}, the inclusion of non-Gaussian gates makes classical simulation challenging. Nevertheless, non-Gaussian gates are necessary for universal quantum computation~\cite{hillmann2020universal,sharma2020characterizing}.

\subsection{Encoding of a layered feed forward architecture}

Classical  feed-forward neural networks (FNN) have layers which can be modeled as such:

\begin{equation}
\mathbf{f}_n(\mathbf{x}) = \sigma(\mathbf{W} \mathbf{x} + \mathbf{b}),
\label{eq:FNN}
\end{equation}
where the $\textbf{x}$ variable stands for the input data, the $\textbf{W}$ and $\textbf{b}$ transformations are respectively the weigh matrix and the bias vector, and $\sigma$ activation function allows to introduce non-linearities in the overall transformation.
As proved by Cybenko and generalized by Hornik~\cite{cybenko1989approximation,hornik1989multilayer,hornik1991approximation}, by setting suitable activation functions $\sigma$, such architecture allows to approximate any measurable function, making NNs universal approximators, later extended to quantum computers \cite{maronese2022quantumactivationfunctionsquantum}.
Through the combinations of Gaussian and non-Gaussian gates, continuous Variable quantum computers are able to create an universal neural network layer similar to the classical one. The goal is pursued through the application of a sequence of gates \cite{Killoran2019}:
\begin{equation}
\label{eq:L}
\mathcal{L} = \Phi \circ D \circ U_2 \circ S \circ U_1,
\end{equation}
where the domain of $\mathcal{L}$ is $\mathbb{R}^N$, representing the $N$-dimensional input space, and the codomain is $\mathbb{R}^N$, representing the $N$-dimensional output space. Here, $N$ denotes the number of input modes, each mode being a real number encoded as a quantum state. All the operations implemented in Eq. \eqref{eq:L} are manipulations of a photonic state, which encode the continuous variables. 
In a fault-tolerant QC era, such operations are aimed to be prompted on a photonic chip \cite{lenzini2018integrated}.

In the Fock representation, each qumode is characterized by position and momentum variables. Such quadratures form a pair per mode that interact with each other, serving as the carriers of information for the neural network. The circuit components of \(\mathcal{L}\) act on these variable carried by qumodes. The operator $U_k = U_k(\boldsymbol{\theta}, \boldsymbol{\phi})$ represents an \(N\)-mode interferometer parameterized by \(\boldsymbol{\theta}\) and \(\boldsymbol{\phi}\), which applies a linear transformation to the input modes. The squeezing operator \(S\) is a single-mode squeezing gate (S-gate) applied to each mode, where the squeezing parameter \(r_i\) acts as a multiplicative factor on the corresponding quadrature. Similarly, the displacement operator \(D\) denotes a single-mode displacement gate (D-gate) that shifts the position and momentum quadratures of each mode by a complex displacement \(\alpha_i\). Finally, the operator \(\Phi\) is a non-Gaussian gate applied to each mode with parameter \(\lambda_i\). A common choice for \(\Phi\) is the Kerr gate (K-gate), which is preferred for its numerical stability-largely due to its diagonal form in the Fock basis.

\begin{figure}
\centering
\subfloat[\label{fig:QNN_layer}]{{\includegraphics[width=0.5\textwidth]{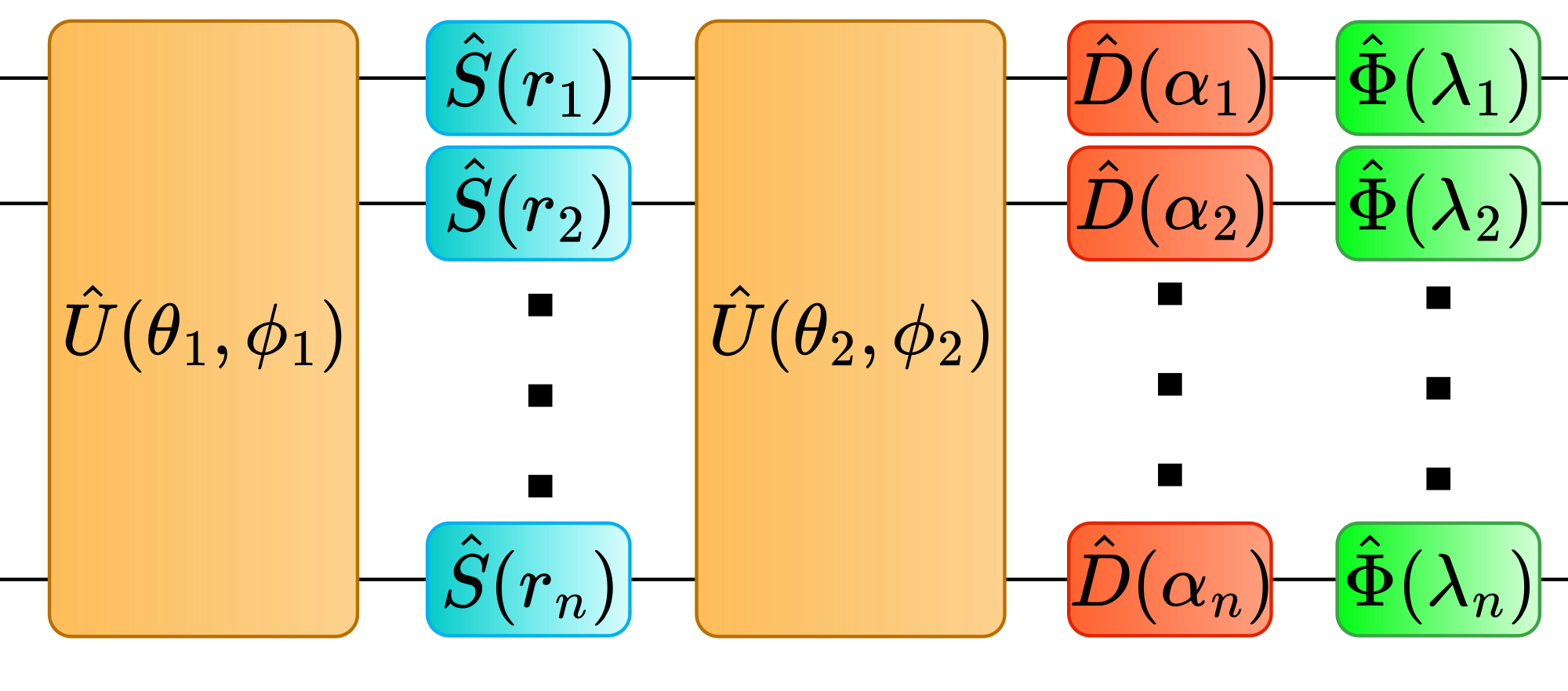} }}
\quad
\subfloat[\label{fig:QLayer_1qumode}]{{\includegraphics[width=0.5\textwidth]{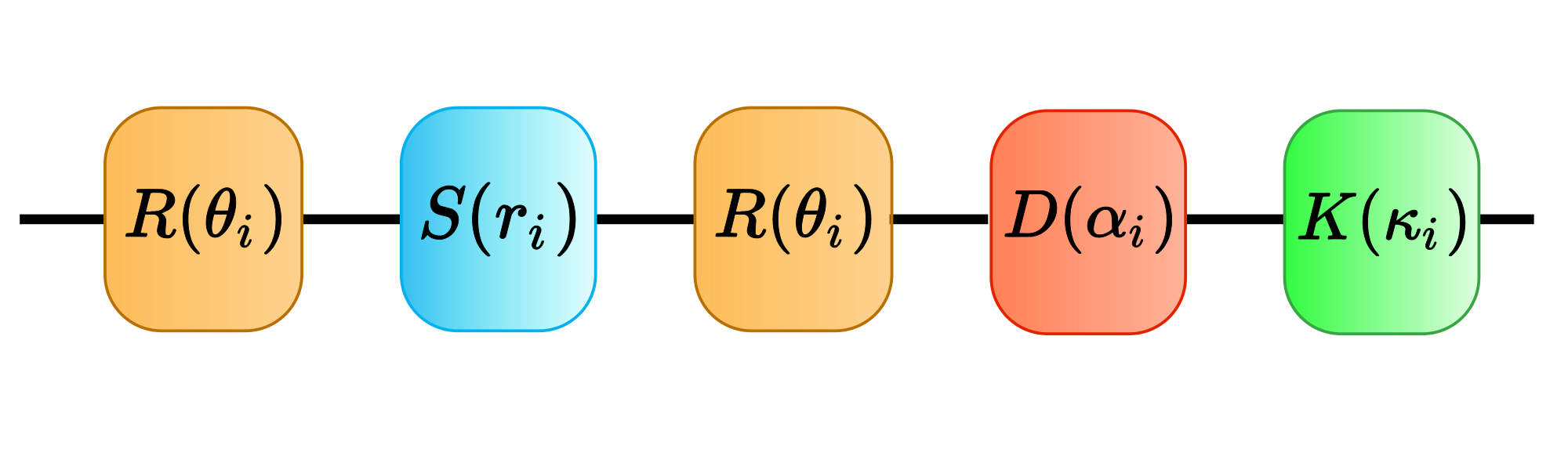}}}
\quad
\subfloat[\label{fig:QLayer_2qumode}]{{\includegraphics[width=0.5\textwidth]{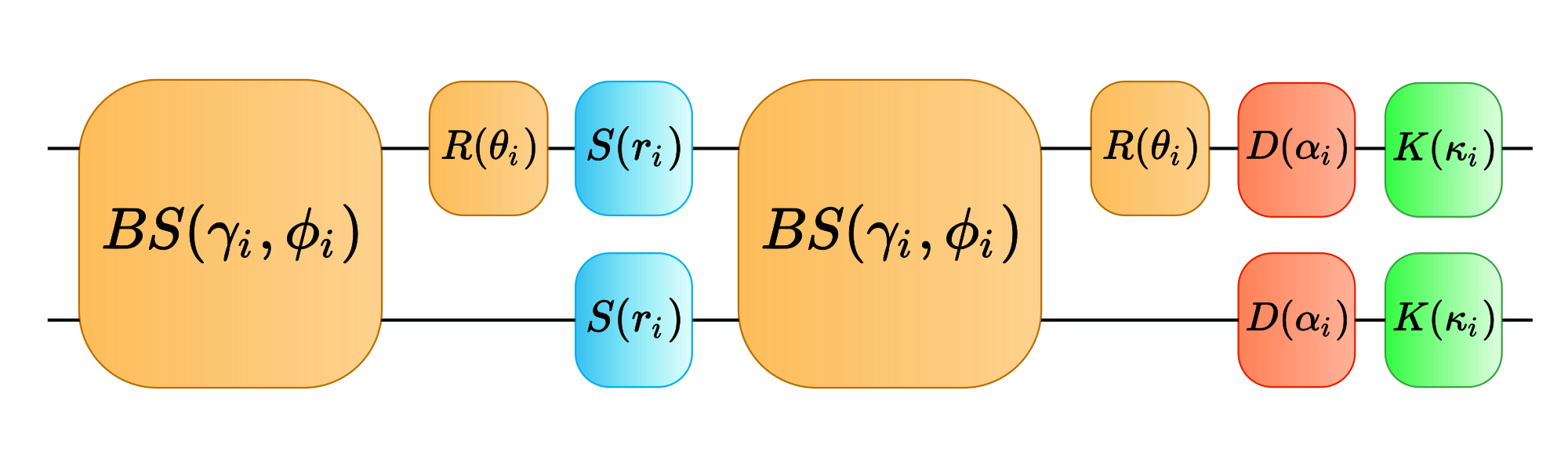} }}
\quad
\caption{(a) Quantum circuit building block for CVQC-based quantum neural network~\cite{strawberryfields2024quantum}. Here $\hat{U}$ represents general n-mode interferometers, $\hat{S}$ represents single-mode squeezing gates, $\hat{D}$ represents single-mode displacement gate and $\hat{\Phi}$ represents any non-gaussian gate (b) \& (c) Quantum Neural Network Layer of for a 1 and 2 qumode set-up respectively. Specifically the 2 modes interferometer is implemented through a Beam Splitter ($BS$) and a phase shifter ($R$), the squeezing ($S$) and displacement ($D$) are single mode and the non-gaussian gate is implement through a Kerr gate ($K$). The 1 qumode configuration correspond to a 2 node fully connected configuration in a classical network but with a higher number of trainable parameters, 5 instead of 6, while the 2 qumode configuration corresponds to a 4 nodes fully connected structure with 12 trainable parameters instead of 20.}
\label{fig:Circuits}
\end{figure}

Therefore the actual structure of the network depends on the number of qumodes, e.g. for a 2 qumode layer the structure is shown in Fig. \ref{fig:QLayer_2qumode}.

\subsection{\label{sub:enc}Encoding of variables}

In the CV formalism, the displacement gate $\hat D(\alpha)$, $\alpha$ being a complex variable, acts as a translation~\cite{Killoran2019}, shifting the $X$ mode by $\Re{\alpha}$ and the $P$ one by $\Im{\alpha}$. As the vacuum state $\ket{0}$ encodes the $0$ value for both the $X$ and $P$ qumodes, a displacement $\hat D(x)$ ($x$ this time being a real parameter), inserted at the beginning of the circuits in Figure \ref{fig:Circuits}, encodes the initial value where the differential equation $\mathcal{F}(x)=0$ is going to be solved. For PDEs, two or more displacement gates are required, depending on the number of variables. 


\subsection{Homodyne detection for quantum measurements}

In the CV framework, the measurements over the quadratures of the electric field can be performed by a technique known as homodyne detection. This technique consists of measuring an observable $\hat Q_\phi$ that is a linear combination of both the observables $\hat X$ and $\hat P$. Nevertheless, in the work we present, as the previous works~\cite{Markidis2022,knudsen2020solving} on QNNs in CV, we are just interested in the $\hat X$ observable. The target function $\textbf{u}_\theta(\textbf{x})$ is yielded in fact by quantum measurements via the homodyne detection:
\begin{equation}
\label{eq:homodyne}
    \textbf{u}_{\boldsymbol \theta}(x) = \bra{\psi_{\boldsymbol \theta}} \hat X \ket{\psi_{\boldsymbol \theta}}
\end{equation}
where $\ket{\psi_{\boldsymbol{\theta}}}$ stands for the quantum state at the end of the transformations induced by the circuit, as in Figure \ref{fig:Circuits}. The family of parameters 
$\boldsymbol{\theta}$ groups all the angles $\theta_i$, $\phi_i$ from the interferometers, the squeezing factor $r_i$ as well as the parameters $\alpha_i$ and $\lambda_i$ for the displacement and the non-linear operations.
For sake of simplicity, henceforth we write $\textbf{u}(x)$ instead of $\textbf{u}_{\boldsymbol{\theta}}(x)$.
Nonetheless, in the scheme we propose, whenever the circuit is composed by multiple waveguides, other output functions can be computed as well via a homodyne detection on a different mode. In the scheme of the 1D Poisson equation, see Figure \ref{fig:Poisson}, we deployed two waveguides to achieve greater expressivity, therefore we adapted such scheme to perform the following measurements:
\begin{equation}
    \begin{cases}
        \textbf{u}(x) = \bra{\Psi_{\boldsymbol \theta}} \hat X_1 \ket{\Psi_{\boldsymbol \theta}} \\
        \textbf{u}_x(x) = \bra{\Psi_{\boldsymbol \theta}} \hat X_2 \ket{\Psi_{\boldsymbol \theta}} \\
    \end{cases}    
\end{equation}
where $\textbf{u}_x(x)$ is a function which should correspond to the $\textbf{u}'(x)$, derivative of $\textbf{u}(x)$, as we detail in Section \ref{sec:1D_Poisson}.

In the scheme for the 2D heat equation, 
the same strategy has been adopted to compute $T$ and $T_x$:
\begin{equation}
    \begin{cases}
        T(x, t) = \bra{\Psi_\theta} \hat X_1 \ket{\Psi_\theta} \\
        T_x(x, t) = \bra{\Psi_\theta} \hat X_2 \ket{\Psi_\theta} \\
    \end{cases}    
\end{equation}

\subsection{Fock space and cut-off truncation}

When simulating the network, we use the Fock space representation, which, upon introducing squeezing gates and displacement gates, becomes an infinite-dimensional space. To efficiently simulate the states, it is necessary to truncate the Fock space at a specific cut-off. Its value has to be carefully chosen before the simulation to achieve an optimal balance between computational efficiency and accurate representation of the manipulated states. We can minimize the cut-off dimension ($n$) by looking at the domain and the predicted co-domain of the equations. By ensuring:
\begin{equation}
 n \gg \max_{x\in\mathcal{D}} x  \land n \gg \max_{x\in\mathcal{D}} u(x) .
\end{equation}

we can grant the cut-off to be sufficient. Moreover, since we are using a variational quantum algorithm (VQA)~\cite{cerezo2021variational}, we can impose the state normalization as one of the components of the loss function, which we explain more in depth in the following Sections.

\subsection{Considerations on the Scaling Complexity of the Problem}
The QNN ansatz for an \(N\)‑qumode system is constructed from \(L\) layers.  Each layer consists of an \(N\)‑mode interferometer \(\hat U\) (decomposed into beam splitters and phase shifters via Clements’ scheme), followed by \(N\) single‑mode squeezing gates \(\hat S\), \(N\) single‑mode displacement gates \(\hat D\), and \(N\) non‑Gaussian (Kerr) gates \(\Phi\) (Fig. \ref{fig:Circuits}).  By Clements’ decomposition, the interferometer alone requires \(O(N^2)\) beam splitters and phase shifters, while the additional Gaussian and Kerr gates contribute \(O(N)\) operations per layer.  Therefore the ansatz cost scales as
\begin{equation}
  G_{\rm total} = G_{\rm ansatz}
                = O\bigl(L\,(N^2 + N)\bigr)
                = O(L\,N^2).
\end{equation}

Regarding the error bound, in classical feed-forward ReLu networks the error is inverse to the size of the network\cite{Yarotsky2017}.
\begin{equation}
    size = O(ln(1/\varepsilon))
\end{equation}

Even though non-Gaussian circuits are known to be universal for continuous‑variable quantum computation \cite{Killoran2019}, as of now there is no proof of a convergence law, connecting the depth of the circuit (L) with the error \(\varepsilon\).
We can only observe a hint of a similar behavior in Figure \ref{fig:Poisson}, i.e. the error scales inversely with the number of layers, both in the classical as in the quantum PINN.

A fully rigorous bound on \(L(\varepsilon)\) for our exact non-Gaussian ansatz remains an open problem.

Therefore the number of quantum operation \textit{ansatz G} is proportional to the squared number of qumodes per layer. Since we need a qumode to encode each variable, the complexity of the circuit will be proportional to the square dimension of the problem (i.e. the number of variables).
\begin{equation}  
    G_{ansatz} = O\bigl(d^2\,L\bigr)
\end{equation}
To draw a conclusion, we assessed the complexity law for the number of operations related with the depth of the circuit $L$ and of the register $N$ (or alternatively $d$). Further studies are required to address how the error $\varepsilon$ does scale by increasing the number of layers $L$.

\subsection{Training the Quantum Neural Network}

To train a quantum neural network, we need to perform backpropagation techniques, i.e. to compute the derivatives of the loss function with respect to the training parameters of the gates. Currently, there is no exact method for obtaining these derivatives in the presence of both Gaussian and non-Gaussian gates~\cite{Schuld_2019}. The best known approach relies on a continuous-variable parameter shift rule, which only works efficiently if the number of non-Gaussian gates grows at most logarithmically with respect to the Gaussian gates \cite{Schuld_2019}. Unfortunately, our universal architecture \cite{Killoran2019} includes one non-Gaussian gate per layer for qumode, violating this condition.

As a result, when training the network directly on quantum hardware, gradient-free optimization techniques have to be employed. During simulation, however, this limitation can be sidestepped by using automatic differentiation methods. A choice made available by Strawberry Fields with a TensorFlow backend, is the automatic Adam optimizer, which has been shown to yield the fastest convergence rates for continuous-variable quantum neural networks~\cite{Killoran2019}.

\section{Quantum Physics-Informed Neural Networks}

Building on the QNN framework described in the previous Section, we now shift our focus on integrating the specialized methods developed for PINNs into our approach.

\subsection{Merging of Quantum Neural Networks and Physics-Informed Architecture}

Quantum Physics-Informed Neural Networks (QPINNs) blend the capabilities of Quantum Neural Networks with the principles of Physics-Informed Neural Networks. 
The network is trained to approximate a target solution function by taking problem-domain inputs and adjusting its parameters to satisfy the underlying physics, as encoded in the governing equations and boundary conditions.
In this framework, the loss function is constructed as a function of the governing differential equations and their associated constraints. By incorporating boundary conditions and other relevant terms into the loss, the QPINN guides the network toward physically accurate solutions. The approximated solution is given by the network, parametrized by the set of gates parameters $\boldsymbol{\theta}$:
\begin{equation}
    \mathbf{u}_{\boldsymbol{\theta}}(\textbf{x}) = \bra{\psi_{\theta}} \hat{X} \ket{\psi_{\theta}},
\end{equation}
where the input to the quantum network is \(\hat D(x)\ket{0}\). For the definition of $\ket{\psi_{\boldsymbol{\theta}}}$, we address the readers to Equation \eqref{eq:homodyne}, while for the encoding through the displacement operator $\hat D(x)$ to Section \ref{sub:enc}.

\subsection{Construction of the loss function}

Let's now turn to the detailed description of the components of the loss function. The loss function is designed to drive the network toward a solution that satisfies the underlying physical laws and boundary conditions. It is composed of several terms each one corresponding to a constraint of the problem. Importantly, one of the terms, called \textit{consistency loss}, represents the core of the method to overcome the issue of nested derivation mentioned above. The loss function is later evaluated at each training epoch. 
Given a general PDE:        
\begin{equation}
\mathcal{N}(\mathbf{u}_{\boldsymbol{\theta}}) = f(\mathbf{x}), \quad \mathbf{x} \in \Omega,
\end{equation}
with boundary conditions:
\begin{equation}
\mathcal{B}(\mathbf{u}_{\boldsymbol{\theta}}) = g(\mathbf{x}), \quad \mathbf{x} \in \partial\Omega.
\end{equation}

Here, \(\mathcal{N}\) is a differential operator acting on the unknown solution \(\mathbf{u}(\mathbf{x})\), \(\Omega\) is the domain, and \(\partial\Omega\) is its boundary. In a physics-informed learning framework, we approximate the solution by a parametric model \(\mathbf{u}_\theta(\mathbf{x})\), in this case via a quantum neural network with parameters \(\theta\).

The training of a QPINN involves minimizing a loss function that incorporates the PDE and boundary conditions as well as other relevant terms. 
Specifically, the loss function includes a physics loss $\mathcal{L}_{\text{PDE}}$, a boundary condition loss $\mathcal{L}_{\text{BC}}$, a trace loss $\mathcal{L}_{\text{trace}}$, an additional term of loss $\mathcal{L}_{\text{extra}}$ and the consistency loss $\mathcal{L}_{\text{consistency}}$, which replaces the need of nested derivation.

The \textit{physics loss} enforces the neural network solution \(\mathbf{u}_\theta\) to satisfy the underlying PDE, by penalizing the deviation of the PDE operator \(\mathcal{N}(\mathbf{u}_\theta)\) from the known forcing term \(f\):
\begin{equation}
\mathcal{L}_{\text{PDE}} = \sum_{\mathbf{x}_i \in \Omega} \left\|\mathcal{N}\bigl(\mathbf{u}_\theta(\mathbf{x}_i)\bigr) - f(\mathbf{x}_i)\right\|^2.
\end{equation}

The \textit{boundary conditions loss}  ensures that the solution \(\mathbf{u}_\theta\) respects the boundary conditions given by the operator \(\mathcal{B}\) and the target function \(g\):
\begin{equation}
\mathcal{L}_{\text{BC}} = \sum_{\mathbf{x}_j \in \partial\Omega} \left\|\mathcal{B}\bigl(\mathbf{u}_\theta(\mathbf{x}_j)\bigr) - g(\mathbf{x}_j)\right\|^2.
\end{equation}

The \textit{trace loss} induces the normalization of the state. Consider the normalized quantum state \(|\psi_{\theta}\rangle\), its density matrix is \(\rho = |\psi_{\theta}\rangle \langle \psi_{\theta}|\). The trace of \(\rho\) should be 1 for proper normalization. Thus, we write:
\begin{equation}
\mathcal{L}_{\text{trace}} = \sum_{i=1}^{N} \bigl(\mathrm{Tr}(|\psi\rangle \langle \psi|)-1\bigr)^2.
\end{equation}

The \textit{additional terms} can include various constraints, regularization terms, or other problem-specific conditions to improve convergence and solution quality:
\begin{equation}
\mathcal{L}_{\text{extra}} = \sum_{k} \lambda_{k} \mathcal{L}_k(\mathbf{u}_\theta),
\end{equation}
where each \(\mathcal{L}_k(\mathbf{u}_\theta)\) is tailored to the problem at hand, weighted by a user-defined hyperparameter $\lambda_{k}$.

The \textit{consistency loss} we introduce consists of a loss component that ensures that the second output of the network corresponds to the derivative of the first output. Instead of relying on nested gradient computations, we directly use the second output as an approximation of the derivative of the first output, which itself approximates the implicit function.

Even though variational quantum circuits rely on parameter shift rule to update their weights, CV computation can efficiently exploit this technique only when the number of non-Gaussian gates is kept logarithmic with respect to the size of the circuit~\cite{schuld2019evaluating}. Such a hurdle prevents to apply the parameter shift rule for the architecture we adopted from Killoran et al., addressing this issue for future works. Numerical difference methods have been tested, proving nevertheless inefficient. The reason for such a failure lies first in the low number of collocation points: for the heat equation, we displaced $10 \times 18$ collocations points, making quite unfeasible to compute a numerical derivative. Moreover, the numerical approach yields incorrect results the more we get close to the boundary of the domain. Instead, 

by enforcing this consistency, we can compute the remaining components of the loss function without requiring nested gradients. 
Specifically, we achieve this by differentiating the second output of the neural network:
\begin{equation}
\mathcal{L}_{\text{consistency}} = \frac{d}{dx} u(x) - u_x(x).
\label{eq:consistency}
\end{equation}
where $u_1(x)$ is the first output of the QNN while $u_2(x)$ is its second output. 

Combining all these loss components into a single total loss function:

\begin{multline}
\mathcal{L}_{\text{total}} =  \lambda_{PDE} \mathcal{L}_{\text{PDE}} + \lambda_{BC}\mathcal{L}_{\text{BC}} + \lambda_{trace} \mathcal{L}_{\text{trace}} + \\
+\lambda_{\text{extra}}\mathcal{L}_{\text{extra}} + \lambda_{consistency }\mathcal{L}_{\text{consistency}} =  \sum_{\ell} \lambda_{\ell} \mathcal{L}_{\ell}  
\end{multline}
where the extra loss term runs in turn  over the index \(\ell\), and all the loss terms are weighted by a suitable  \(\lambda_{\ell}\) hyper-parameters which are their corresponding weighting coefficients.

\begin{figure}[!htbp]
    \centering
    \includegraphics[width=\linewidth]{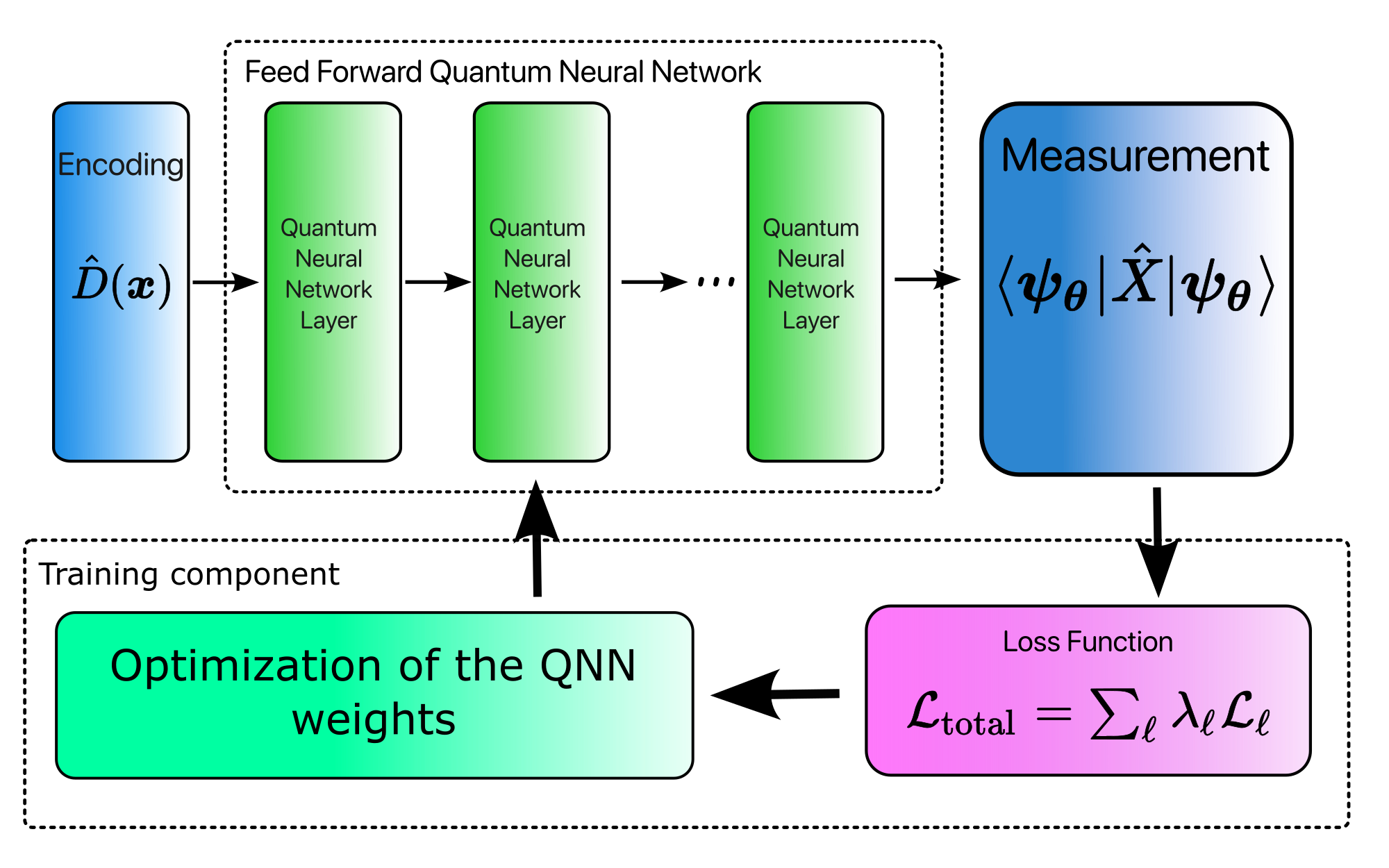}
    \caption{QPINN overall structure. The encoding is implemented by a set of displacement gates $\hat{D(\boldsymbol{x})}$ while output is obtain through the homodyne detection of the position quadrature of the final state $\hat{X}$. A classical computer then calculates the loss function a carries optimization by updating the weights of the parametrized QNN layers.}
    \label{fig:QPINN_struct}
\end{figure}

After each loss evaluation, the parameters $\boldsymbol{\theta}$ of the quantum neural network are updated using the Adam optimizer with automatic differentiation. Iterating this optimization process leads the QPINN to converge toward an accurate target solution \(\mathbf{u}(\mathbf{x})\).
The structure of the QPINN is represented in Fig. \ref{fig:QPINN_struct}.
\subsection{Power consumption of a photonic device}

An interesting benchmark between classical networks and quantum circuits concerns the power consumption of the two devices. In the first place, we focus on the paramount component of all optical devices, the mesh of Mach-Zehnder interferometers (MZIs). MZIs are designed as two $50:50$ beam splitters, the second taking as input the two output beams of the first one. In the mid, a phase-shifter alters the optical length of one of the two paths~\cite{clements2016optimal}. The rotations $\hat U(\theta_i, \phi_i)$ in Figure \ref{fig:QNN_layer} are indeed implemented by a mesh of MZIs.
Building and designing an efficient MZI mesh is a key-feature to unlock fault-tolerant photonic quantum computing~\cite{ranjan2023experimental,smith2022universal,taballione2021universal}. An MZI mesh allows the implementation of multiplications of unitary matrices in a highly parallelized way~\cite{reck1994experimental,clements2016optimal}, thus emerging as a potential competitor against classical devices such as GPUs. As proved by Clements~\cite{clements2016optimal}, a universal multimode interferometer can be built up by blocks of MZIs preceded by a phase-shifter. This scheme requires two phase shifters as power consumptive components, the 50:50 beam-splitters being passive. On silicon devices, phase modulations can be induced shifting the temperature in the phase-shifting regions. These devices take the name of thermo-optics phase shifters (TOPS), with an average power consumption of $2.56mW$ for a $\pi$ phase shift~\cite{xiao2023recent}. 
Since the number of interferometers scale as $O(n^2/2)$ in both Reck's and Clements' schemes~\cite{reck1994experimental,clements2016optimal}, the number of phase shifters required amounts to $O(2n^2)$.
As a rough evaluation, we could say that the energy consumption follows the above scheme. Side effects, such as cross-talk between phase-shifters, have to be considered, but it strictly depends on the material and design of the devices, addressing this characterization for a more specialized research. Adopting these parameters, a $4 \times 4$ matrix multiplication would require nearly $80mW$, while for a $100 \times 100$ matrix the consumption is close to $6.4W$. A classical GPU, instead, has an idle load of $50W$. Even though the scaling of the quantum devices cannot actually compete with classical GPUs, it still turns to be competitive for edge applications when dealing with small-sized matrices.
Nevertheless, an alternative to the MZI mesh is the loop architecture. Within this frame, the light travels along a spiral circuit, with two loops (an inner one and an outer one) interfaced by a programmable Mach-Zehnder interferometer.
Such an architecture could in principle drastically reduce power consumption, since the power consumption would refer to this sole component. Still, in principle, the deeper the algorithm to be run, the longer the execution time, depending on the specific algorithm under consideration.

Other gates, such as displacement or squeezing, can be easily implemented on the vacuum state, but up to now there is lack of standard techniques about how to apply them in the mid of the circuit. Just for instance, some schemes for universal computation have been proposed~\cite{takeda2017universal} but tailor-suited for loop architectures. For such a reason, until an integrated chip implementing a layer of a neural network is not available, we consider a benchmark between a classical network and a photonic circuit to be premature.

\section{\label{sec:lintrouction}Results }

Quantum Physics-Informed Neural Networks (QPINNs) leverage Quantum Processing Units (QPUs) 
to operate the neural network and compute its gradients. Subsequently, a classical computer 
is employed to update the weights parameters of the neural network by using a user-defined optimizer.

In our implementation, we utilize \textit{Strawberry Fields}, a Python library developed by 
Xanadu \cite{strawberryfields2024quantum}, to simulate the quantum circuits. Meanwhile, 
\textit{TensorFlow} is employed to perform automatic differentiation via its 
\texttt{tf.GradientTape} method \cite{tensorflow2015-whitepaper}.

Achieving reliable convergence in QPINNs generally requires substantial fine-tuning. We refer the reader to Appendix~A for details on the specific hyperparameters used and the rationale behind them Tables \ref{tab:hyper_Poisson}, \ref{tab:hyper_heat}.  For all the simulations we run, we adopt Adam optimizer. Nevertheless, a comparative benchmark between different common optimizer is carried in Appendix \ref{app:optimizers}.

We proceed in two steps. First, we introduce a new architecture designed to overcome existing limitations associated with nested derivative operations, a capability that had previously induced significant challenges. Next, building upon this architectural innovation, we demonstrate the successful solution of a partial differential equation (PDE), specifically the heat equation, using this improved QPINN framework.

\subsection{Solution of an ODE: case of 1D Poisson equation} \label{sec:1D_Poisson}

\begin{figure*}[!htbp]
    \centering
    \begin{minipage}{\textwidth}
        \centering
        \subfloat[]{%
            \includegraphics[width=\textwidth]{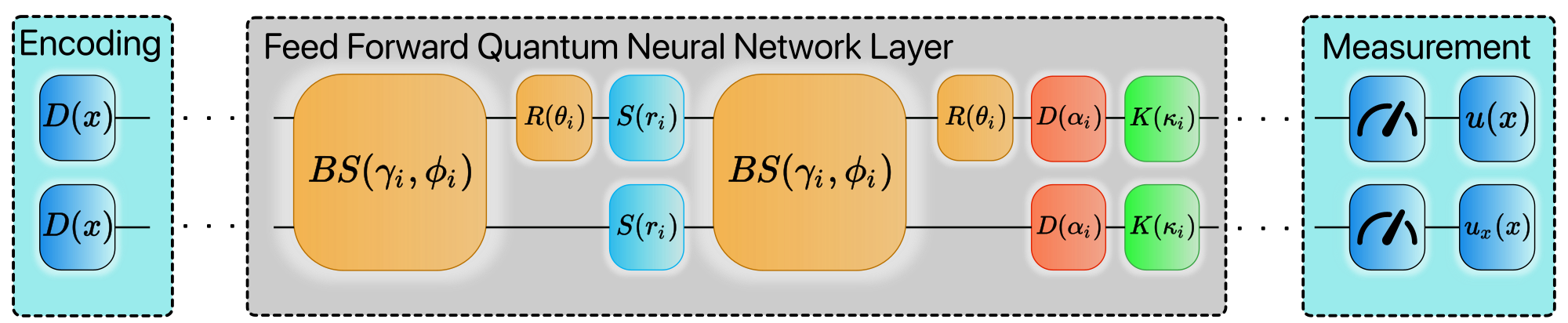}
            \label{fig:Poisson_a}
        }\\[1ex]
        \subfloat[]{%
            \includegraphics[width=0.8\textwidth]{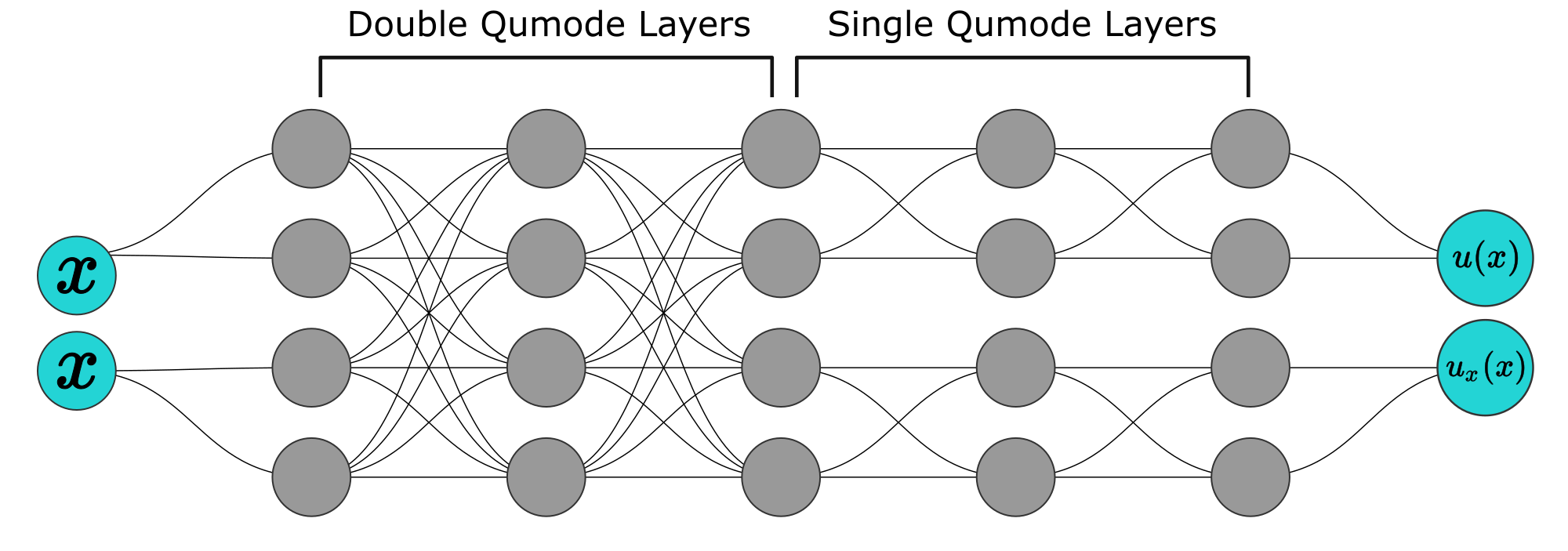}
            \label{fig:Poisson_b}
        }
        \vspace{1ex} 
        \caption{(a) Schematic network setup used to solve the 1D Poisson equation. The input \(x\) is encoded as qumode displacement.
        The network applies linear and non-linear transformation in the form of gaussian ($BS$, $R$, $S$, and $D$) and non-gaussian ($K$) gates to achieve the desired output. The first output of the QNN is measured and used to estimate the implicit function $u(x)$. The second one is used to estimate the first derivative of the implicit function \( u'(x) \).
        (b) Classical equivalent of the actual QNN used to solve the 1D Poisson equation. The layer in (a) corresponds to a 4 nodes layer in a classical network.}
        \label{fig:Poisson}
    \end{minipage}
\end{figure*}

With this first case, a neural network is built to solve ordinary differential equations (ODEs) with an architecture designed to avoid gradient nesting while being able to implement higher order derivation. We used a multi-output configuration in which the second output of the network converges towards the derivative of the first output (Fig. \ref{fig:Poisson}). To showcase this architecture we solve the 1D Poisson equation with a sinusoidal term (Eq. \ref{eq:1D_Poisson}).

\begin{equation}
    \label{eq:1D_Poisson}
    u''(x) + \sin(4x) = 0,
\end{equation}
with boundary conditions and domain

\begin{equation}
    u(0) = u\left(\frac{\pi}{2}\right) = 0, \qquad
    x \in \left[0, \frac{\pi}{2}\right].
\end{equation}
To solve this equation we set up the network as in Figure \ref{fig:Poisson}.
%


The loss function components are illustrated in Table \ref{tab:loss_Poisson_weights}.
\begin{table}[ht]
\renewcommand{\arraystretch}{1.3}    
\setlength{\tabcolsep}{12pt}          
\centering
\begin{tabular}{lll}
\toprule
\textbf{Loss} & \textbf{Expression} & \textbf{Weights} ($\lambda$) \\
\midrule
\(\mathcal{L}_{\mathrm{PDE}}\) & 
\(\left(\frac{\partial u_x(x)}{\partial x} + \sin(4x)\right)^2\) 
& 25\% \\
[6pt]
\(\mathcal{L}_{\mathrm{BC}}\) & 
\(u(0)^2+u\left(\frac{\pi}{2}\right)^2\)
& 25\% \\
[6pt]
\(\mathcal{L}_{\mathrm{consistency}}\) & 
\(\Bigl(\tfrac{\partial u}{\partial x} - u_x\Bigr)^2\)
& 25\% \\
[6pt]
\(\mathcal{L}_{\mathrm{trace}}\) & 
\(\Bigl|\langle \psi \mid \psi \rangle - 1 \Bigr|^2\)
& 25\% \\
\midrule
\(\mathcal{L}_{\mathrm{tot}}\) & 
\(\displaystyle \sum_{\ell} \lambda_{\ell}\,\mathcal{L}_{\ell}\)
& 100\% \\
\bottomrule
\end{tabular}
\caption{Loss function components used in the 1D Poisson equation problem with corresponding weights. No particular configurations of the weights was found to be particularly better towards the resolution of the equation.}
\label{tab:loss_Poisson_weights}
\end{table}

\begin{figure}[!htbp]
    \centering
    \includegraphics[width=\linewidth]{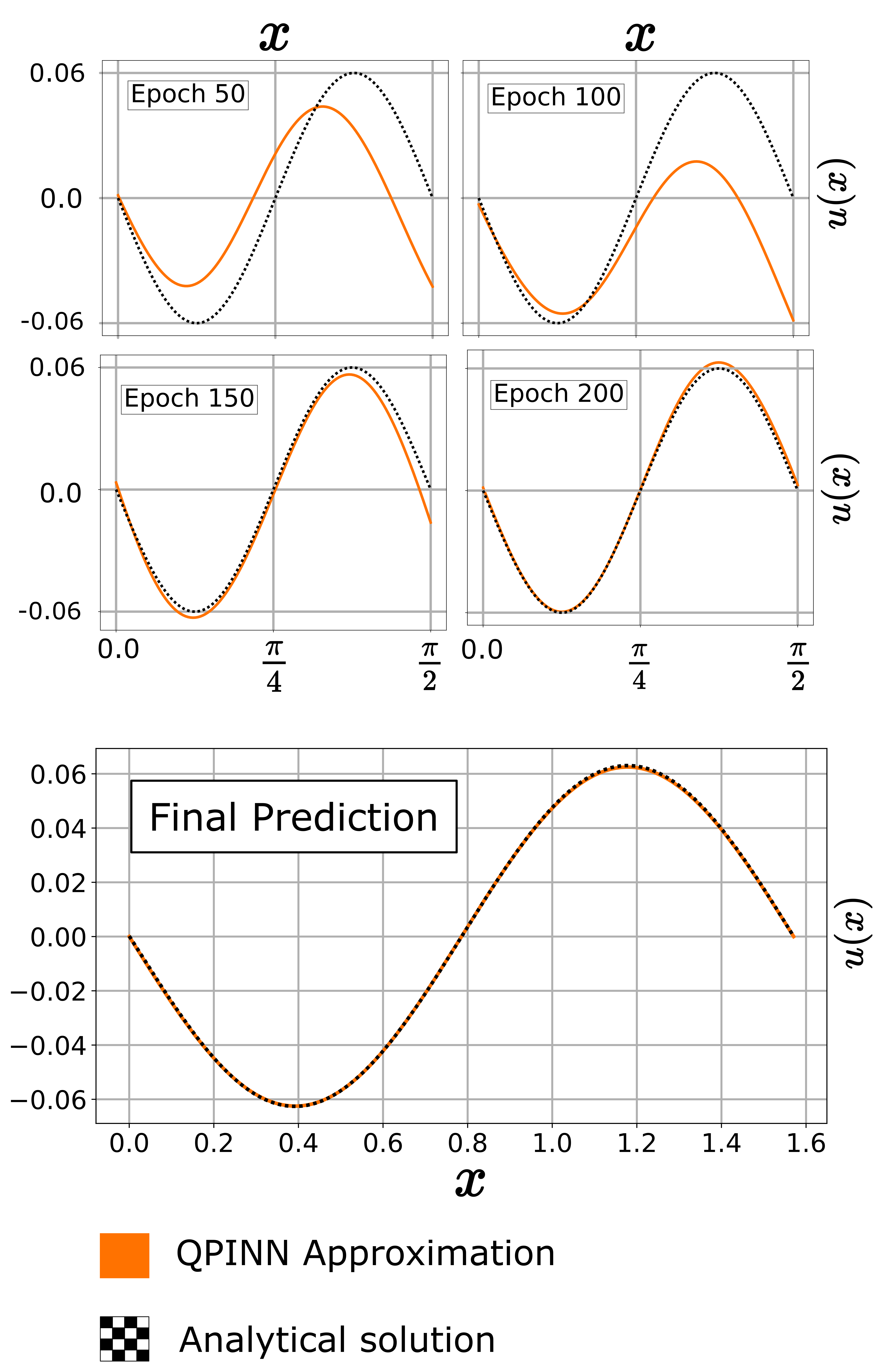}
    \caption{Training progress to solve the case of ODE represented by the 1D Poisson equation without gradient nesting while being able to implement higher order derivation.
    Top: Evolution of the prediction of the QPINN after 50, 100, 150 and 200 epochs of training. Bottom: the best result obtained by the method corresponding to   RMSE $= 1.09\times 10^{-4}$ }
    \label{fig:1D_Poisson_evo}
\end{figure}

\begin{figure}[!htbp]
    \centering
    \includegraphics[width=\linewidth]{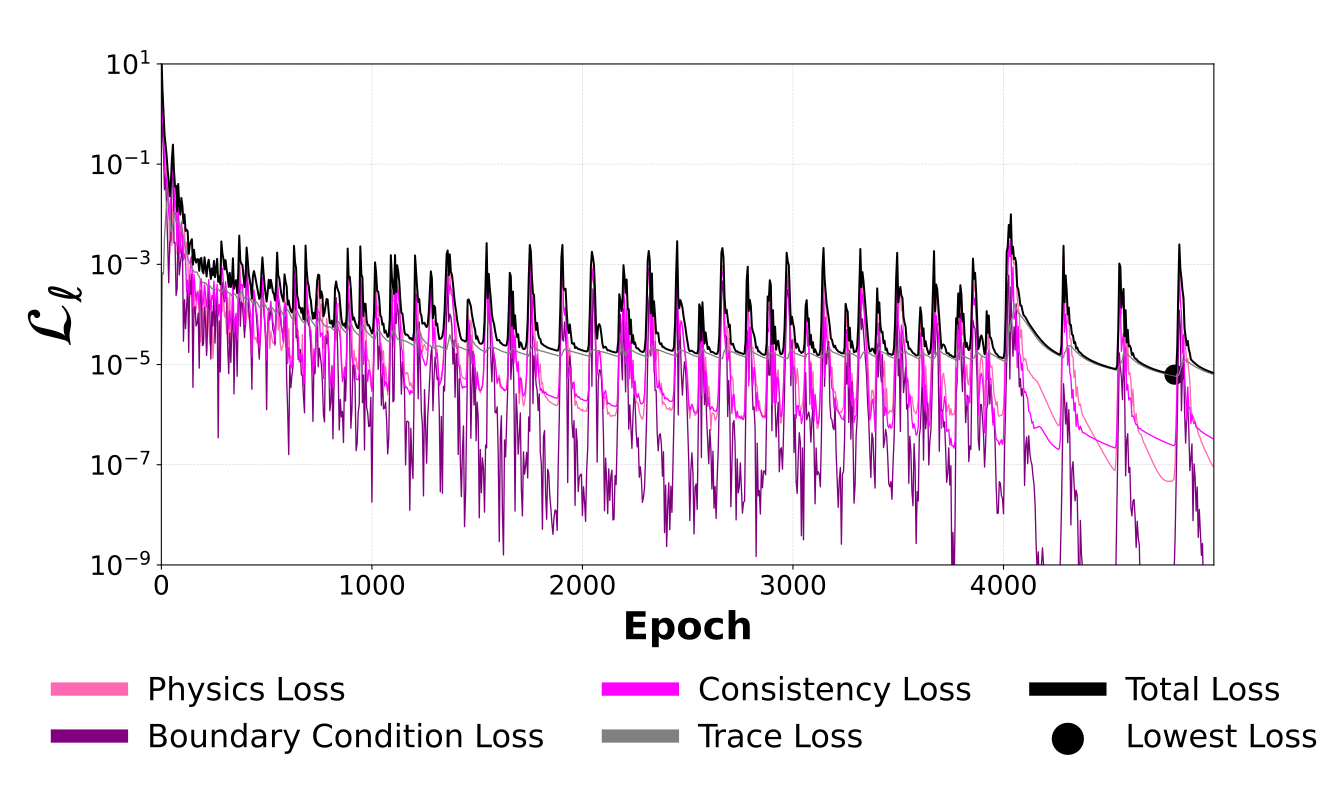}
    \caption{Different components of the loss function during the training of the neural network for the solution of the 1D Poisson equation. A black dot is used to indicate the epoch with the lowest loss value. The values are plotted every 20 epochs to make the lines clearer. The optimization is able to achieve at its best a lowest loss given by the RMSE $= 1.09\times 10^{-4}$ as measured in Eq. \ref{eq:Poisson_NMSE}. Similarly to what happens with some other machine learning methods, the method can show instabilities, requiring to memorize the best network weights during the training when performances are improved.}
    \label{fig:loss_1D_Poission}
\end{figure}
To ensure that the domain is correctly represented and each point in the domain is fitted, the points evaluated are randomly drawn from the domain, according to sobol sequence \cite{berg2018unified}.

As illustrated in Fig.~\ref{fig:1D_Poisson_evo}, the QPINN method solves this equation efficiently. The Root Mean Squared Error (RMSE) of the final solution, measuring its accuracy, is defined as
\[
    \mathrm{RMSE} = \sqrt{\frac{1}{N}\sum_{i=1}^{N} \bigl(y_i - \hat{y}_i\bigr)^2},
\]
which for the best result of the 1D Poisson equation solved by our method is:
\begin{equation}
    \mathrm{RMSE} = 1.09 \times 10^{-4}.
    \label{eq:Poisson_RMSE}
\end{equation}
This metric reflects the typical magnitude of the point‑wise error in the same units as the solution.

We also report the Normalized Mean Squared Error (NMSE), which measures the MSE normalized by the magnitude of the solution:
\[
    \mathrm{NMSE} = \frac{\frac{1}{N}\sum_{i=1}^{N} \bigl(y_i - \hat{y}_i\bigr)^2}
                       {\frac{1}{N}\sum_{i=1}^{N} y_i^2}.
\]
For our computed solution, this yields:
\begin{equation}
    \mathrm{NMSE} = 6.08 \times 10^{-6}.
    \label{eq:Poisson_NMSE}
\end{equation}
This dimensionless ratio indicates what fraction of the solution’s total power is accounted for by the error.

These low values indicate the convergence of the method and validate its effectiveness in solving higher-order differential equations, as shown in Figure  \ref{fig:loss_1D_Poission}.
For the sake of completeness and in order to make the Reader aware of the behavior of the method in practice, we report there the full record of a training process, which may show sudden instabilities, as often happens during training in other artificial intelligence methods, such as the deep reinforcement learning which sometimes suffers of sudden loss of performances \cite{corli2021solving,corli2023casting,moro2022goal}. In these cases, the solution consists of recording the weight parameters returning at the minimum loss achieved.

We performed a quantitative comparison of QPINNs and classical PINNs to establish a fair benchmark of their respective capabilities. To ensure comparability, both models were configured to have a similar total number of trainable parameters. Although QPINNs were executed on classical hardware,making a perfect apples-to-apples comparison impossible, we believe matching architecture sizes provides a reasonable proxy. Accordingly, we constructed fully connected feed-forward neural networks with varying numbers of layers so that each classical PINN has approximately the same parameter count as its QPINN counterpart. The exact layer configurations and total number of parameters are given in Table \ref{tab:params_qpinn_vs_pinn}.

\begin{figure}
    \centering
    \includegraphics[width=\linewidth]{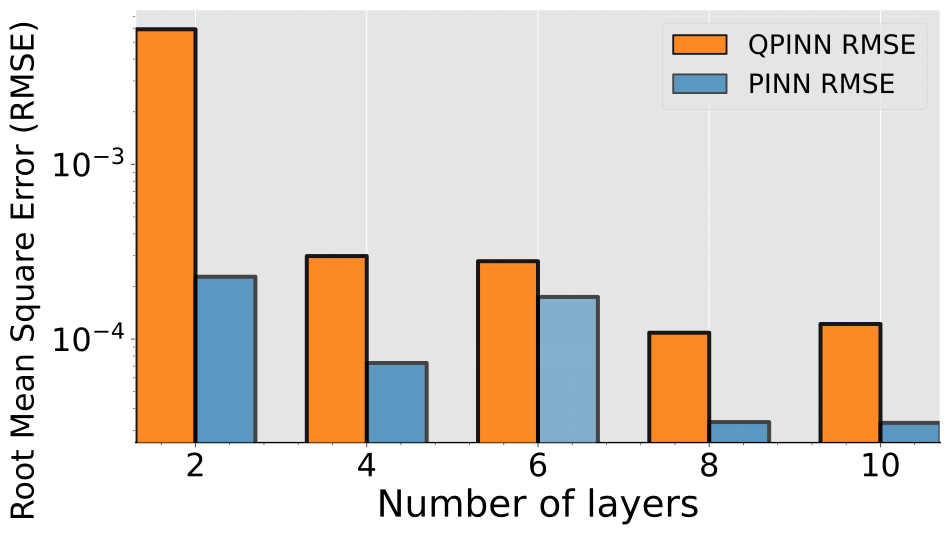}
    \caption{  Comparison of QPINN and PINN on the Poisson equation with similar architectures. The QPINN schematic (Fig. \ref{fig:Poisson_a}) employs 22 trainable weights every two layers, while the classical PINN uses three neurons per layer, yielding a comparable total number of trainable parameters (see Table \ref{tab:params_qpinn_vs_pinn}). In some cases the classical PINN converges more rapidly to the solution and provides a finer final precision. Nevertheless the two methods have comparable precision which could be improved in both  cases through simulation's hyper-parameters fine tuning.}
    \label{fig:rmse_vs_layers}
\end{figure}

\begin{table}[ht]
\centering
\begin{tabular}{c c c}
\hline
\textbf{\# of hidden layers} & \textbf{QPINN} & \textbf{PINN} \\
\hline
2  & 22  & 22  \\
4  & 44  & 46  \\
6  & 66  & 70  \\
8 & 88  & 94  \\
10 & 110 & 118 \\
\hline
\end{tabular}
\caption{  Number of trainable weights used in the comparison between QPINN and PINN in the Poisson equation solution. }
\label{tab:params_qpinn_vs_pinn}

\end{table}

\subsection{Solution of a PDE: case of the Heat Equation}

The method involving the consistency loss is now applied to PDEs. As application case, we consider the heat equation, which describes the temporal evolution of the temperature distribution in a one-dimensional medium. It is given by Eq. \ref{eq:heat}:
\begin{equation}
    \label{eq:heat}
    \frac{\partial T}{\partial t} = \alpha_d \frac{\partial^2 T}{\partial x^2}, \quad x \in [-\frac{\pi}{2}, \frac{\pi}{2}], \; t \geq 0,
\end{equation}
where \(T(x,t)\) is the temperature at position \(x\) and time \(t\), and $\alpha_d$  is the thermal diffusivity.
In this equation, each variable has a physical interpretation, but we deliberately omit units of measurement. Our goal is to demonstrate the method's effectiveness in solving second-order, linear, parabolic partial differential equations, exemplified here by the heat equation (Eq.~\ref{eq:heat}). Importantly, the method remains invariant under rescaling of the domain, therefore changing the setting of the boundary conditions does not affect the quality of the solution.
The boundary conditions are:
\begin{equation}
    T(-\frac{\pi}{2}, t) = T(\frac{\pi}{2}, t) = 0, \quad t \geq 0,
\end{equation}
We set $\alpha_d = 0.30$ and the initial condition:
\begin{equation} 
    T(x, 0) =  \frac{1}{2}\exp\left(-\frac{\left(x + \frac{\pi}{8}\right)^2}{2\sigma^2}\right) = T_0( x), \quad \sigma^2 = 0.2.
\end{equation}
We simulate the evolution from $t = 0$ to $t = 0.5$.

\begin{figure*}[!htbp]
    \centering
    \begin{minipage}{\textwidth}
        \centering
        \subfloat[]{%
            \includegraphics[width=\textwidth]{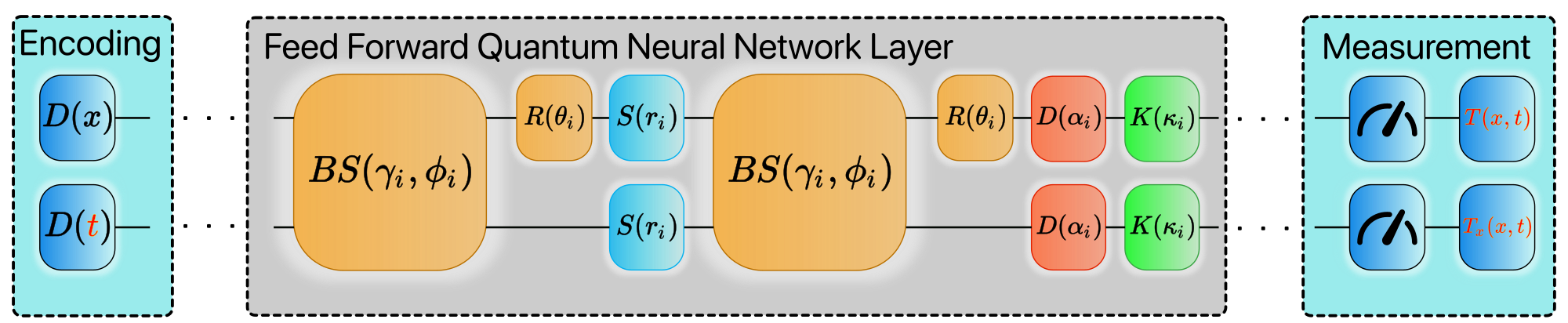}
            \label{fig:Heat_a}
        }\\[1ex]
        \subfloat[]{%
            \includegraphics[width=0.8\textwidth]{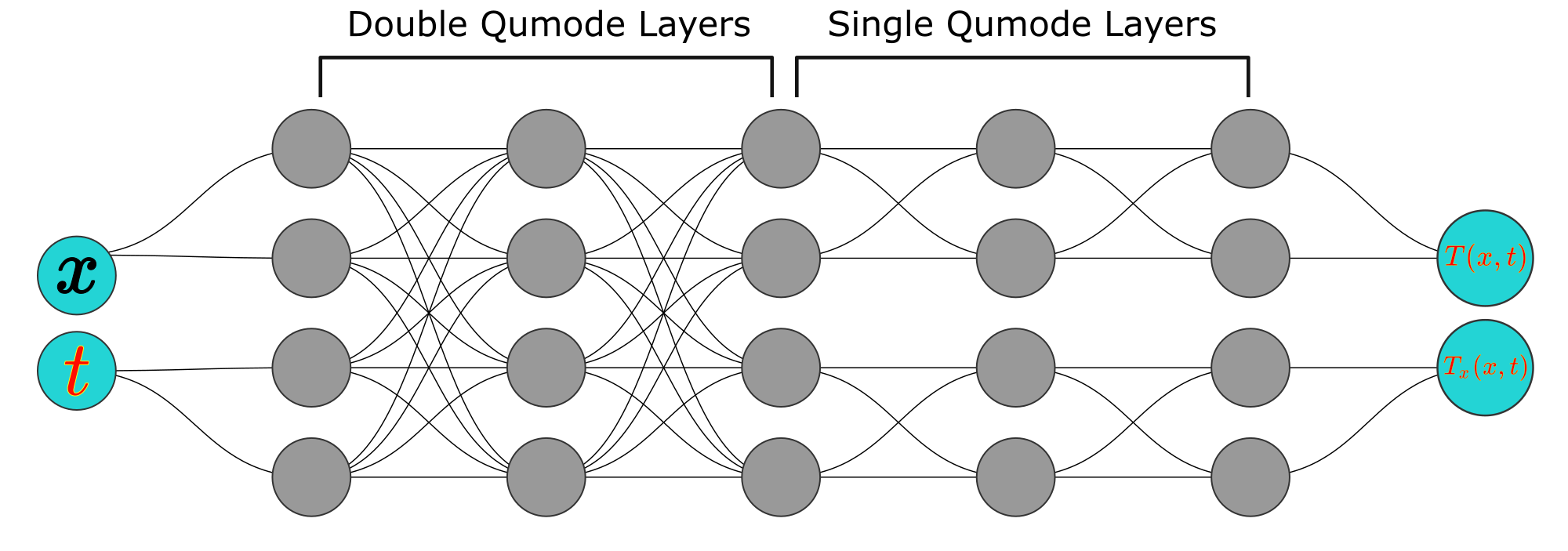}
            \label{fig:Heat_b}
        }
        \vspace{1ex} 
        \caption{(a) Schematic network set-up used for solving the Heat equation. In this case, differently from Fig. \ref{fig:Poisson} the inputs \(x\) and \(t\) are encoded as qumode displacements. The network applies linear and non-linear transformation in the form of gaussian (BS, R, S and D) and non-gaussian (K) gates to achieve the desired output. The network outputs are measured and give two values: \(T(x,t)\), an approximation of the desired solution \(u(x,t)\), and \(T_x(x,t)\), an approximation of \(\frac{\partial u}{\partial x}\). For the heat equation we need only 2 outputs since we do not need to calculate \(\frac{\partial^2 T(x,t)}{\partial t^2}\); therefore, we only need to estimate the first derivative with respect to position.
        (b) Classical equivalent of the actual QNN used to solve the 1D Poisson equation. The layer in (a) corresponds to a 4 nodes layer in a classical network.}
        \label{fig:Heat}
    \end{minipage}
\end{figure*}

In the QPINN, both \(x\) and \(t\) are encoded into a two-qumode quantum neural network (QNN) by applying displacement gates on each qumode to represent these inputs. The network architecture consists of a two-layer, two-qumode section followed by a two-layer, single-qumode section (see Fig.~\ref{fig:Heat}). This structure enables us to approximate both the heat equation's solution \(T(x,t)\) and its spatial derivative \(\frac{\partial T}{\partial x}\), thanks to the consistency loss (Eq. \ref{eq:consistency}). We only need this second output for this specific equation because the equation does not require us to compute the second derivative with respect to time, therefore requiring only two outputs.

We denote the two outputs of the network as \(T(x,t)\) and \(T_x(x,t)\). By automatic differentiation of these outputs, we obtain the necessary derivatives \(\frac{\partial T}{\partial x}\), \(\frac{\partial T}{\partial t}\), and \(\frac{\partial T_x}{\partial x}\). These derivatives are then used to form the loss function components, ensuring that the learned functions satisfy the loss functions (Table \ref{tab:loss_IC_weights}).

\begin{table}[ht]
\renewcommand{\arraystretch}{1.3}    
\setlength{\tabcolsep}{12pt}          
\centering
\begin{tabular}{lll}
\toprule
\textbf{Loss} & \textbf{Expression} & \textbf{Weights} ($\lambda$) \\
\midrule
\(\mathcal{L}_{\mathrm{PDE}}\) & 
\( \left( \frac{\partial T}{\partial t} - \alpha_d \frac{\partial T_x}{\partial x}\right)^2\)
& 10\% \\
[6pt]
\(\mathcal{L}_{\mathrm{IC}}\) & 
\(\left(T(x,0)-T_0(x)\right)^2\)
& 60\% \\
[6pt]
\(\mathcal{L}_{\mathrm{BC}}\) & 
\(T\left(-\frac{\pi}{2}\right)^2+T\left(\frac{\pi}{2}\right)^2\)
& 10\% \\
[6pt]
\(\mathcal{L}_{\mathrm{consistency}}\) & 
\(\Bigl(\tfrac{\partial T}{\partial x} - T_x\Bigr)^2\)
& 10\% \\
[6pt]
\(\mathcal{L}_{\mathrm{trace}}\) & 
\(\Bigl|\langle \psi \mid \psi \rangle - 1 \Bigr|^2\)
& 10\% \\
\midrule
\(\mathcal{L}_{\mathrm{tot}}\) & 
\(\displaystyle \sum_{\ell} \lambda_{\ell}\,\mathcal{L}_{\ell}\)
& 100\% \\
\bottomrule
\end{tabular}
\caption{Loss function components with corresponding weights: 60\% for \(\mathcal{L}_{\mathrm{IC}}\) and 10\% for the others. These weights were chosen bacause of the particular physics if the problem, i.e to prevent the network from ending up in the local minima consisting in the null solution ($T(x,t) = 0$) which would satisfy the differential equation but not the initial conditions. }
\label{tab:loss_IC_weights}
\end{table}
To ensure that the domain is correctly represented and each point in the domain is fitted, the evaluated points  are randomly drawn from the domain, according to sobol squence \cite{berg2018unified}.

To reduce computational cost and accelerate training the network can be "pre-trained", i.e. we can train the network in the beginning just by using the initial condition loss which does not require derivative calculations.

The network is then trained for 300 epochs with 258 points distributed on the $T(x,0)$ which greatly reduces computational times.
Afterwards the network is trained on the equation full domain with all of its loss function components.
Cosine annealing with warm restarts is employed to prevent the network from converging to global minima (\cite{loshchilov2017sgdrstochasticgradientdescent}).

\begin{figure*}[ht]
    \centering
    \includegraphics[width=1\linewidth]{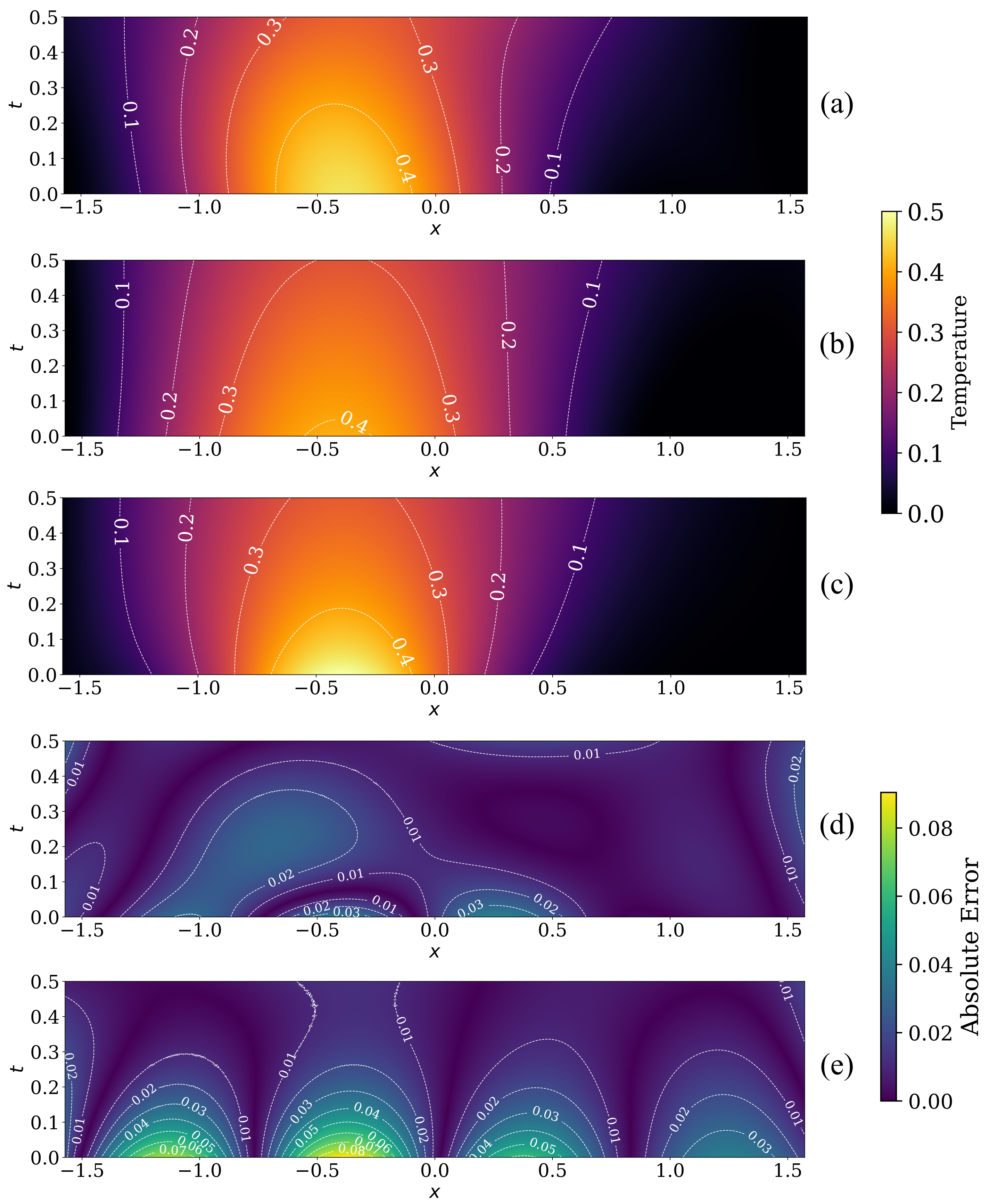}
    \caption{  Comparison of the best Quantum PINN (QPINN) and classical PINN against a Runge–Kutta (RK) reference. (a, b) Temperature predictions from QPINN and PINN over space x and time t (light = hot), with dashed contours marking isotherms. (c) RK solver solution. (d, e) Absolute error of QPINN and PINN versus RK. Both networks use 44 trainable weights, were pre-trained for 300 epochs on the initial-condition loss, then trained 1000 epochs on the full PINN loss. Although run on a simulator, QPINN yields slightly lower error than its classical counterpart, highlighting its competitiveness for future quantum hardware. }
    \label{fig:tavola_heat}
\end{figure*}

\begin{figure}
    \centering
    \includegraphics[width=\linewidth]{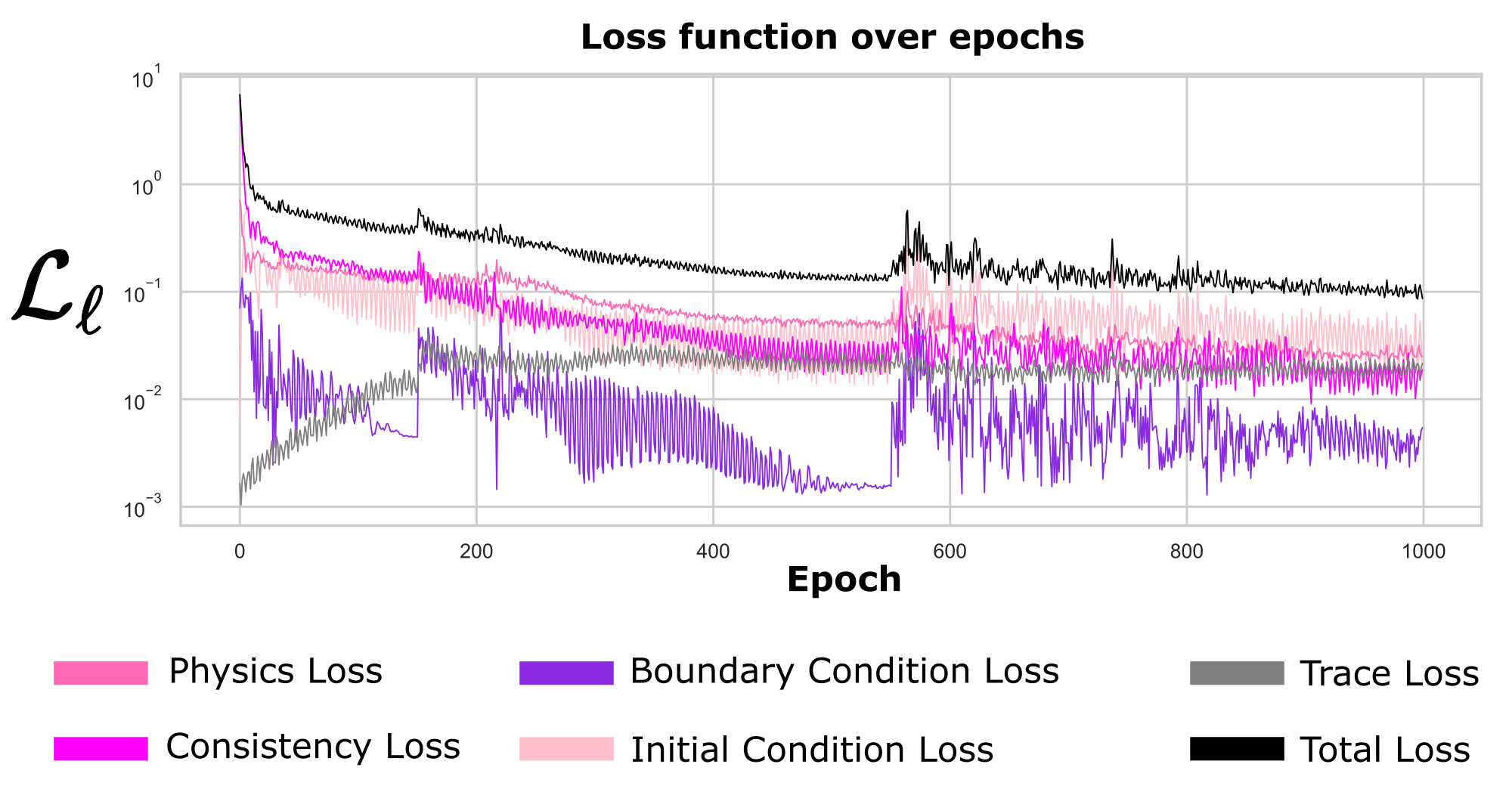}
    \caption{Loss function components values during the training of the QPINN used to estimate the solution of the Heat Equation (Eq. \ref{fig:tavola_heat}). The model is trained for 1000 epoch and a steadly descent in the loss function is appreciatable. We can see that in the end of the training the loss function reaches a values of $10^{-1}$ which has to be divided by 100 (total weights of the loss function) to obatian an estimate of the final precision of the method  $10^{-3}$}
    \label{fig:heat_loss}
\end{figure}

By minimizing the total loss, the QPINN learns a solution that is consistent with the PDE, boundary and initial conditions, and the quantum model constraints.
After 1000 training epochs, the network successfully converges to an accurate solution of the equation (Fig. \ref{fig:tavola_heat}). The loss function decreases steadily throughout the training process, except for a spike at epoch 500, ultimately reaching a value on the order of \(10^{-1}\)  (Fig. \ref{fig:heat_loss}).

We compare our solution with a classical physics-informed neural network (PINN). As in the Poisson’s equation case, we use the total number of trainable parameters as our benchmark. The classical PINN is a fully-connected feed-forward network with a single hidden layer of 11 neurons, for a total of 44 trainable weights. To assess accuracy, both networks are evaluated against a reference solution computed via SciPy’s adaptive, embedded Dormand–Prince Runge–Kutta integrator (RK45) \cite{butcher2008,hairer1993, dormand1980family}, by measuring the pointwise absolute error (see Figure \ref{fig:tavola_heat}). Both models achieve comparable precision: the larger residual in our network likely stems from the classical constraints of its simulation, while the classical PINN may be under-expressive given its smaller parameter count and fewer training epochs. Table \ref{tab:metrics_heat} summarizes these results, showing that our network attains a slightly lower error. However, because there are many tunable hyperparameters in each model (learning rate, number of epochs, network depth and width, etc.) and because simulating our approach is computationally intensive, a truly quantitative comparison, and any claim of an intrinsic advantage, remains difficult.

{
  \renewcommand{\arraystretch}{2.5}  
  \begin{table}[ht]
    \centering
    \begin{tabular}{lccc}
      \toprule
      \textbf{Metric} & \textbf{Formula} 
                      & \textbf{QPINN} & \textbf{PINN} \\
      \midrule
      
      RMSE              & $\displaystyle \sqrt{\frac{1}{N}\sum_{i=1}^N (y_i - \hat y_i)^2}$
                        & $1.24\times10^{-2}$ & $2.09\times10^{-2}$ \\ 
      MAE               & $\displaystyle \frac{1}{N}\sum_{i=1}^N \lvert y_i - \hat y_i\rvert$
                        & $9.63\times10^{-3}$ & $1.48\times10^{-2}$ \\ 
      $L_\infty$ Error  & $\displaystyle \max_{1\le i\le N}\lvert y_i - \hat y_i\rvert$
                        & $3.93\times10^{-2}$ & $9.04\times10^{-2}$ \\
      NMSE              & $\frac{\sum_{i=1}^N (y_i - \hat y_i)^2}{\sum_{i=1}^N y_i^2}$
                        & $3.17\times10^{-3}$ & $9.06\times10^{-3}$ \\
      \bottomrule
    \end{tabular}
    \caption{ Evaluation metrics of QPINNs and PINNs for the solution of the 1D Heat Equation (see Figure \ref{fig:tavola_heat}). The quantum network slightly outperforms the classical counterpart.}
    \label{tab:metrics_heat}
  \end{table}
}

Nevertheless, this result demonstrates the capability of QPINN in solving PDEs but suggests also that further improvements can further  enhance the method's precision. Further details about the simulation constraints and overheads are discussed in the Appendix.

\section{Realistic simulation based on photonic quantum hardware processor}

In this Section, we assess the effect of noise on QPINNs. In order to characterize such a noise, we rely on the X8 quantum photonic processor provided by Xanadu company~\cite{xanaduX8}. First, we characterize the noise affecting the device, mostly due to photon losses. Afterwards, we emulate the noise on the circuit through an appropriate gate representation.
Eventually, in order to prove the robustness of the variational algorithm with respect to photon losses, we showcase the capability of QPINNs in error mitigation.

\subsection{Modelling}

We refer to the work by Ranjan~\cite{ranjan2023experimental} about the characterization of the X8 device, a reconfigurable 8-qumodes photonic processor developed by Xanadu, able to implement Gaussian Boson Sampling algorithms. There, the authors showcase the number of photons detected after generating squeezed light states. While the interferometers are switched to act as the identity on the circuit, the initial states are set as two-mode squeezed vacuum (TMSV) states, which can be described as 
\begin{equation}
    \ket{TMSV(z)}_{1,2} = \hat S_{1,2}(z) \ket{0} = \exp(z \hat a_1 \hat a_2 - z^* \hat a_1^\dagger \hat a_2^\dagger) \ket{0}
\end{equation}
The factor $z$ is a complex number, which can be described in polar terms as $z=r e^{i\gamma}$.
In that experiment, $\gamma=0$ and $r=1$,. The number of counted photons per mode on the X8 device is accounted as well. Among all the possible sources of noise, they conclude that photon losses play a paramount role.

When characterizing the effects of photon losses on ladder operators and thus conjugated observables, we introduce noisy channels, which can be simulated in Strawberry Fields library through an opportune loss gate $\hat L(T)$~\cite{strawberryfields2024quantum}:
\begin{equation}
    \begin{cases}
    \hat L^\dagger(T) \hat a_i \hat L(T) = \sqrt{T} \hat a_i + \sqrt{1-T} \hat a_{i,e} \\
    \hat L^\dagger(T) \hat a^\dagger \hat L(T) = \sqrt{T} \hat a^\dagger + \sqrt{1-T} \hat a_{i,e}^\dagger
    \end{cases}
\end{equation}
Here, $\hat a_{i,e}$ stands for the external mode -- the environment -- the processor is exchanging photons with, resulting in the optical loss of photons in the computational process. The canonical observables are consequently transformed as follows:
\begin{equation}
    \begin{cases}
        \hat q_i \to \sqrt{T} \hat q_i + \sqrt{1-T} \hat q_{i,e} \\
        \hat p_i \to \sqrt{T} \hat p_i + \sqrt{1-T} \hat p_{i,e}
    \end{cases}
\end{equation}
Here, $T$ represents the losses suffered by the considered
mode~\cite{qi2018linear}. To characterized the overall photon loss suffered by a single mode -- and thus the $T$ factor -- we start from the number of photons counted on the X8 device. This operation can be described as
\begin{equation}
    \langle n_i \rangle = \bra{\psi} \hat n_i \ket{\psi} = 
    \frac{1}{2}\bra{\psi} [\hat q_i^2 + \hat p_i^2 - 1] \ket{\psi}
\end{equation}
with $\ket{\psi} = \hat L(T) \ket{TMSV(r)}$, and $\hat S_{i,j}(r)$ (i.e. the operator to create the TMSV state on the vacuum) can be decomposed into $\hat{BS}(\pi/4) \hat S_i(r) \hat S_j(-r)$. From this consideration, we can act over the canonical variables $\hat q$ and $\hat p$ in the phase space:
\begin{equation}
\label{eq:Lintro}
    \begin{split}
    \frac{1}{2}\bra{0} \hat S^\dagger_{i,j}(r) \hat L_i^\dagger(T_i) [\hat q_i^2 + \hat p_i^2 - 1] \hat L_i(T_i) \hat S_{i,j}(r) \ket{0} = \\
    \frac{1}{2}\bra{0} \hat S^\dagger_{i,j}(r) [(\sqrt{T_i} \hat q_i + \sqrt{1-T_i} \hat q_{i,e})^2 + \\
    + (\hat p_i \sqrt{T_i} + \sqrt{1-T_i} \hat p_{i,e})^2 - 1] \hat S_{i,j}(r) \ket{0}
    \end{split}
\end{equation}
Afterwards, acknowledging that
\begin{subequations}
\begin{align}
    \hat{BS}_{i,j}^\dagger(\pi/4) \hat q_i \hat{BS}_{i,j}(\pi/4) = (\hat q_i \pm \hat q_j)/\sqrt{2} \\
    \hat{BS}_{i,j}^\dagger(\pi/4) \hat p_i \hat{BS}_{i,j}(\pi/4) = (\hat p_i \pm \hat p_j)/\sqrt{2} \\
    \hat S^\dagger(r) \hat q \hat S(r) = e^{-r} \hat q \\
    \hat S^\dagger(r) \hat p \hat S(r) = e^{r} \hat p
\end{align}
\end{subequations}
we achieve
\begin{equation}
\begin{split}
    \frac{1}{2}\bra{0} \left[\frac{(e^{-r} \hat q_i \sqrt{T_i} \pm e^{r} \hat q_j \sqrt{T_i} + \sqrt{1-T_i} \hat q_{i,e})^2}{2} + \right. \\
    + \left. \frac{(e^{r}\hat p_i \sqrt{T_i} \pm e^{-r} \hat p_j \sqrt{T_i} + \sqrt{1-T_i} \hat p_{i,e})^2}{2} - 1\right] \ket{0}
\end{split}
\end{equation}
Since $\bra{0} \hat q \ket{0} = \bra{0} \hat p \ket{0} = 0$ and $\bra{0} \hat q^2 \ket{0} = \bra{0} \hat p^2 \ket{0} = 1/2$, for both the internal and external modes, only the squared terms survive (not the mixed ones), and we get
\begin{equation}
\begin{split}
    \frac{1}{2}\left(T_i \frac{\cosh{2r}}{2} + \frac{1-T_i}{2} + T_i \frac{\cosh{2r}}{2} + \frac{1-T_i}{2} - 1 \right) = \\
    T_i \frac{\cosh(2r) -1}{2} = T_i \sinh^2(r)
\end{split}
\end{equation}
When the system is ideal (i.e. $T=1$), the above result is consistent with literature~\cite{li2021experimental}. Eventually, it is possible to assess the noise in the single mode by the following formulation:
\begin{equation}
\label{eq:T_i}
    T_i = \frac{\langle n_i \rangle}{\sinh^2(r)}
\end{equation}

\subsection{Device characterization}

\begin{figure}
\centering
\subfloat[\label{subfig:layer_loss_a}]{{\includegraphics[width=\linewidth]{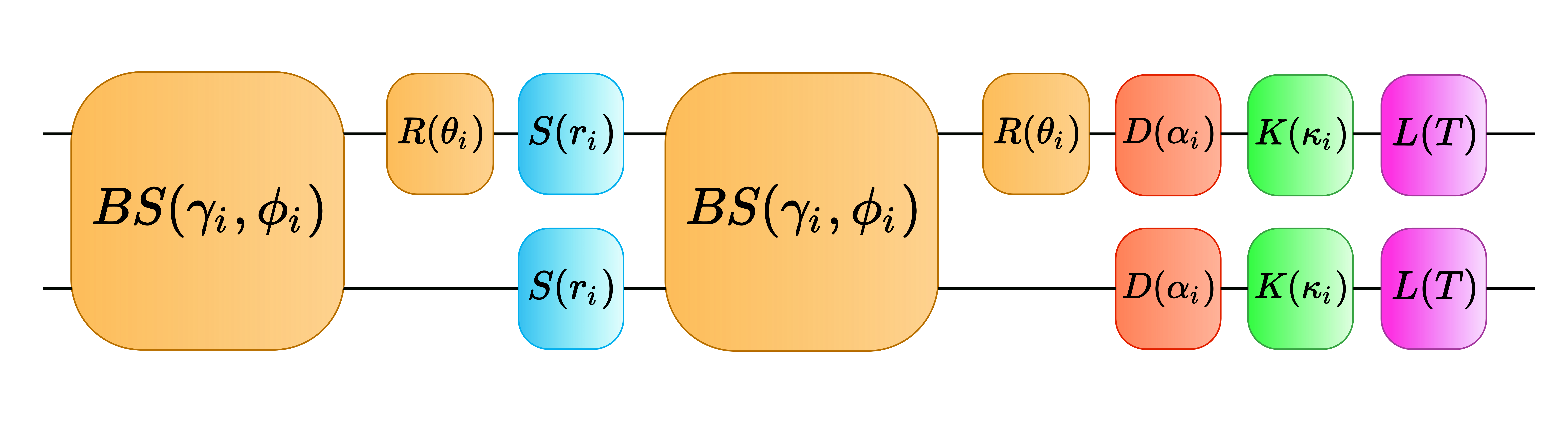} }}
\quad
\subfloat[\label{subfig:layer_loss_b}]{{\includegraphics[width=\linewidth]{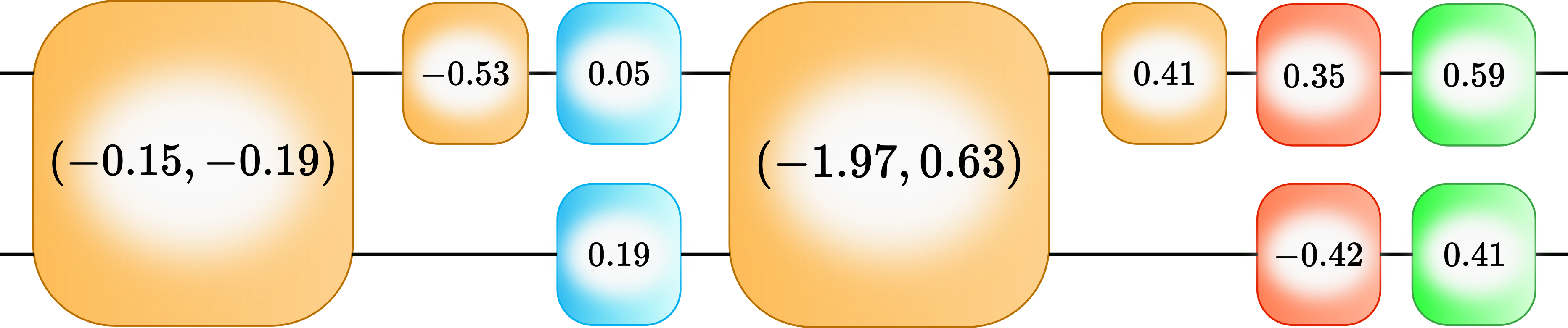}}}
\quad
\subfloat[\label{subfig:layer_loss_c}]{{\includegraphics[width=\linewidth]{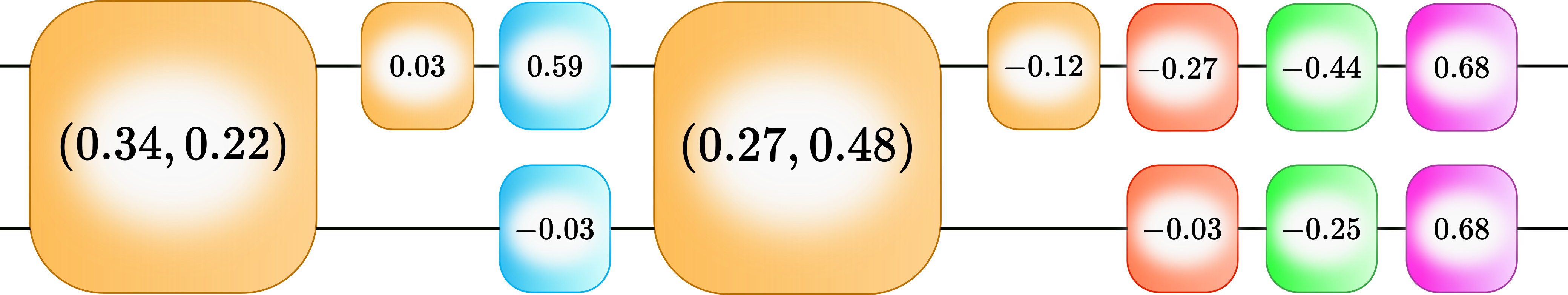} }}
\quad
\caption{ Representation of a noisy layer of a QPINN. (a) Two qumode layer of a QNN with addition of optical losses. The $\hat L(T)$ gates (in violet) are attached after the logic gates of the ideal layer, previously shown in Figure \ref{fig:QLayer_2qumode}. (b) The Figure reports the arguments of the gates applied for a noiseless and optimal simulation. (c) The Figure reports the re-parametrization of the previous layer with addition of noise, showcasing the adaptability of the quantum network.}
\label{fig:loss_layer}
\end{figure}

We now detail our measurement campaign carried on X8 device. When dealing with QPINNs, this processor does not provide the gates we have described since now, required to run QPINNs (see Section \ref{sec:architecture} and Figure \ref{fig:Circuits}). Indeed, this device provides a rigid mesh of reconfigurable interferometers and a source of TMSV states as inputs. Nonetheless, we can interpret X8 as a QPINN layer, therefore the characterization of this device can provide a realistic model for the losses. To apply such a model, we add to our layer a sequence of gates introducing an attenuation factor, whose values follow our experiments. In Figure \ref{fig:loss_layer}, we propose this model of noisy layer, where, compared to the ideal layer, see Figure \ref{fig:QLayer_2qumode}, we introduce an amount of optical losses per channel due to the $\hat L(T_i)$ gate, already introduced in Equation \eqref{eq:Lintro}, as the last step of the layer.

We characterize the attenuation factor $T_i$ by sampling the number of photons on each channel $i=1,...,8$ and comparing the mean average of counts $\langle n_i \rangle$ to the ideal distribution. Therefore, by plugging the values of $\langle n_i \rangle$ in Equation \eqref{eq:T_i},  we achieve the final value of each $T_i$.
All the experiments are run by tuning $r=1$, and the interferometers on null values (i.e. acting as the identity). The outcome of our experiments is reported in Figure \ref{fig:photon_count}. For each channel, we have counted the average $\langle n \rangle_m$, number of measured photons on the detector, compared to the ideal value $\langle n \rangle_{ideal} = \sinh^2(1) \simeq 1.38$. The total amount of shots is equal to 30000.
We then obtain the value of $T_i$ for each channel $i$ by employing Equation \eqref{eq:T_i}. Moreover, we compare the ideal probability distribution of photon counting, computed by a classical emulation, to the data gathered from our experiments.

\subsection{Error mitigation techniques for QPINNs}

Even though NISQ devices are still immature to perform proper error mitigation techniques, VQ algorithms prove to be resilient, leveraging on backpropagation and parameter updating to contrast noise effects, Figure \ref{fig:loss_layer}~\cite{dangwal2025variational,khanal2023evaluating}. Such a strategy proves to be effective mostly when dealing with systemic errors rather than stochastic ones, since the formers can be predicted and mitigated with better accuracy. Therefore, in the first case, adaptive algorithms, such as QPINNs, can learn the error and counterbalance it.

\begin{figure*}[t]
    \centering
    \includegraphics[width=\textwidth]{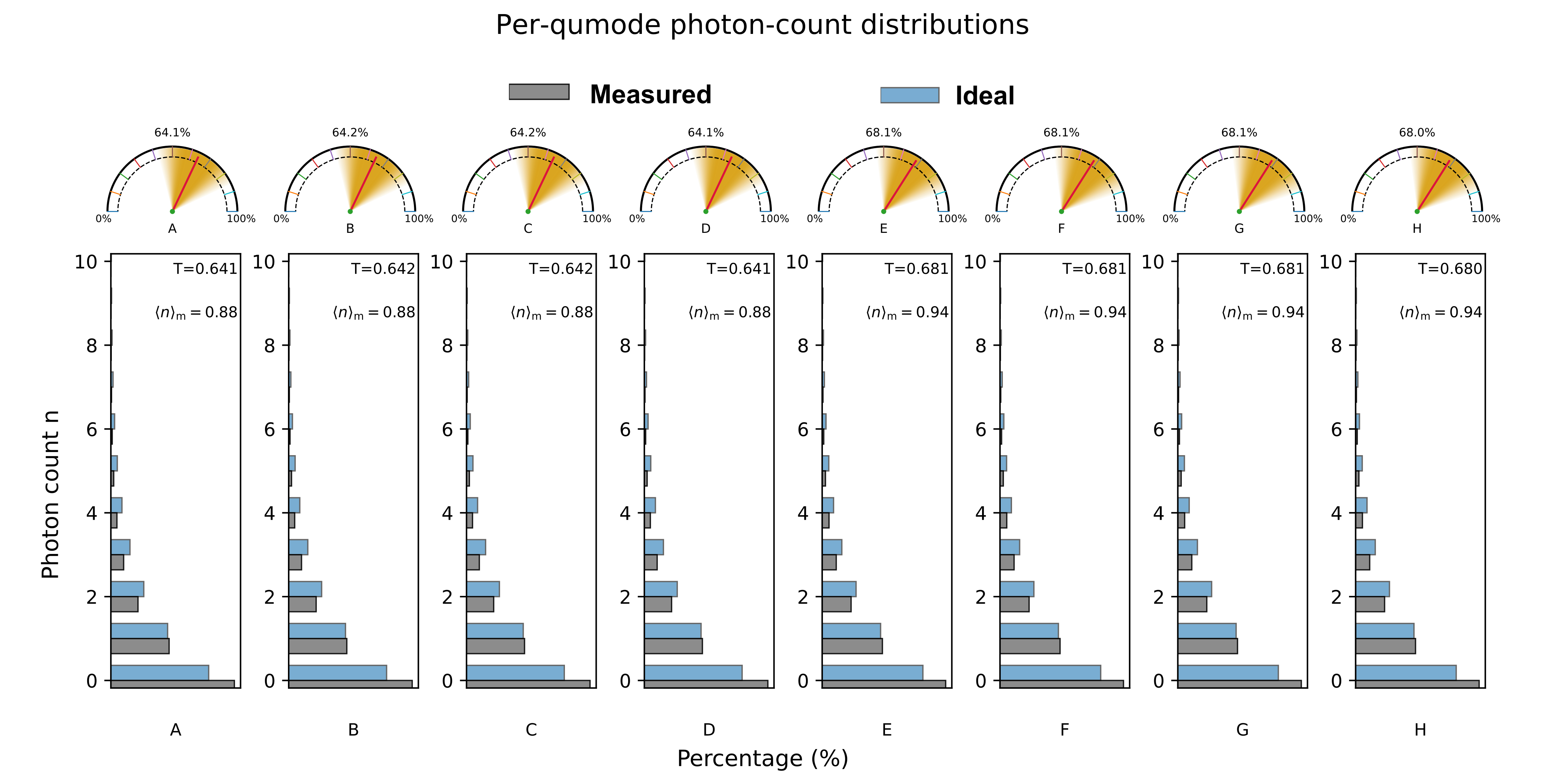}
    \caption{Experimental results from the measurement campaign on X8. Each column represents a channel (from A to H), where the gauges on top display the mean transmittance $T$ (crimson needle) and its standard deviation (ocher sector), while the histogram below the probability distribution of photon counting (in grey) and their ideal counterpart (in cyan). The total number of shots is equal to 30000.}
    \label{fig:photon_count}
\end{figure*}
Optical losses are subsumed in this class of systemic noise, as highlighted in the characterization of X8 processor both from ref. ~\cite{ranjan2023experimental} and our experiments. Such a hypothesis is verified in Figure \ref{fig:photon_count}, where the variance on the photon count per mode is bounded in a small range. Since the channel attenuation $T_i$ can be derived from the photon count, as stated in Equation \eqref{eq:T_i}, these values can be estimated with good accuracy when simulating QPINNs on classical computers. To account such a noise effect, each time the noise is introduced in the QPINN layer as in Figure \ref{fig:loss_layer}, we sample the attenuation $T_i$ from the distribution obtained by the experimental data from X8, see Figure \ref{fig:photon_count}.
When tackling the Poisson 1D equation, since two qumode suffice, we choose to sample $T_{a}$ and $T_{b}$ from the channels with higher average values of photon counts $\{n_i\}_{i=1}^8$. Further studies and techniques on error mitigation for QPINNs -- and optical QNNs in general -- would require additional characterizations of target devices, capable of installing proper Gaussian and non-Gaussian operations arranged as in Figure \ref{fig:Circuits}, which actually X8 is not able to.

\subsection{Numerical results from noisy simulations}

\begin{figure}
    \centering
    \includegraphics[width=\linewidth]{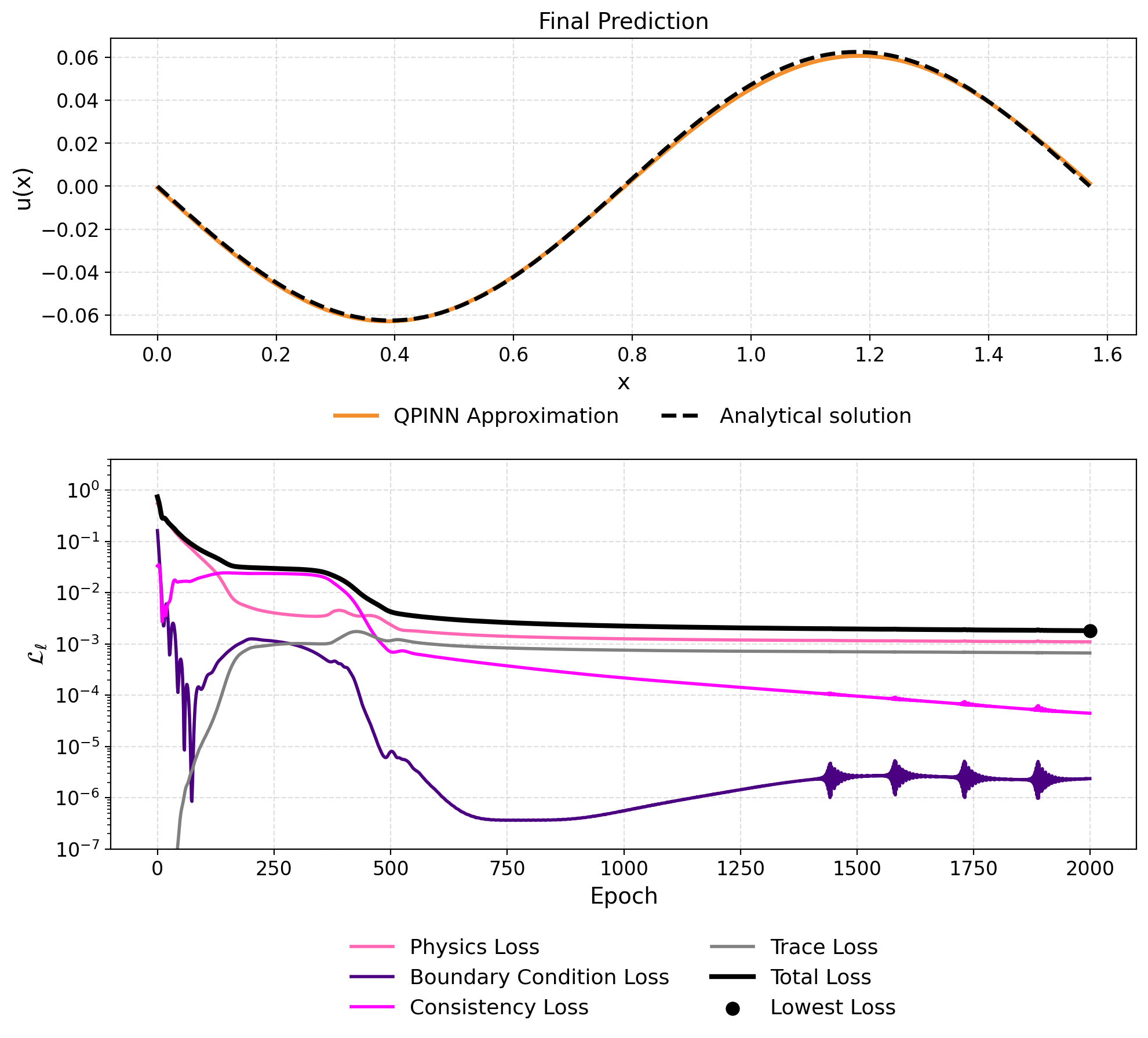}
    \caption{  We show the solution of a 2 layer QPINN used to solve the Poisson equation. The RMSE is $1.065 \times 10^{-2}$ and the NMSE is $5.831 \times 10^{-4}$. The parameters are clearly able to adapt to overcome the problem posed by the loss of photons}
    \label{fig:lossy_sim}
\end{figure}

Here we run a numerical simulation with the loss model mentioned above to confirm the robustness of the method when noise is taken into consideration. We run the QPINN with a two qumode configuration to solve the Poisson equation. The hyperparameters used are the same used in Table \ref{tab:loss_Poisson_weights} except  three of them, namely number of total layers which is two,the number of epochs which is 2000 and the cutoff dimension which is increased to 20 to make sure the network has sufficient expressivity. In the simulation we set a fixed transmittance $T$ per layer to $T=0.68$, assuming that the number of shots averages out the loss to this specific value like in the experiments in Figure \ref{fig:photon_count}. We show in Figure \ref{fig:lossy_sim} that the QPINN can accurately solve the differential equation even in the case of a loss of photons. The  NMSE is $5.831 \times 10^{-4}$ and the RMSE
is $1.065 \times 10^{-3}$ showcasing no deterioration compared to the noiseless solution reported in Figure \ref{fig:rmse_vs_layers}. We therefore conclude that this algorithm is robust against a realistic amout of noise.

\section{Conclusions}

We have demonstrated the capability of quantum physics-informed neural networks to solve partial differential equations (PDEs). Although our experiments were conducted using a quantum simulator, which imposes limitations such as restricting the Fock space to a predefined cutoff dimension and necessitating the rescaling of the equation's domain, such constraints are expected to diminish when transitioning to actual quantum hardware. 

On real quantum devices, the inherent normalization of quantum states eliminates the need for trace loss, and the absence of a fixed cutoff dimension allows for more accurate representations without domain rescaling.
0
However, deploying QPINNs on quantum hardware introduces its own set of challenges that may require modifications to the neural network architecture. One significant bottleneck is the current inability to compute exact derivatives on quantum hardware, which hampers the effectiveness of gradient-based optimization methods essential for training neural networks. While gradient-free optimization techniques can be employed, one should carefully consider their convergence speed and accuracy compared to methods based on exact derivatives.

Despite these hurdles, the potential of QPINNs remains promising. Continued research and development are crucial to address these challenges and fully leverage quantum computing for solving PDEs. 
We have addressed experimental noise of a photonic quantum device to simulate QPINNs in realistic noise environment.
QPINNs prove to be resilient against systemic noise such as optical losses. In order to make QPINNs deployable, further studies are needed to characterize the implementation and corresponding optical losses for any specific hardware architecture.

Future applications may involve integrating many bodies Schr{\"o}dinger equations along with data provided by quantum sensing techniques.

With further advancements, QPINNs have the potential to surpass traditional Physics-Informed Neural Networks (PINNs) in both efficiency and performance.

\section{Acknowledgements}

The authors also acknowledge support from the Qxtreme project funded by the Partenariato Esteso FAIR (grant N. J33C22002830006). GP and EP gratefully thank Thales Alenia Space Italia for supporting a PhD grant.
The authors thank Marco Sampietro and Giorgio Ferrari for the fruitful discussion about the power consumption of optical devices.  EP gratefully thanks Xanadu for having granted access to X8 hardware.
The authors thank Dario Chemoli for the support in calculating the photon distribution in ideal conditions.

\appendix
\section{Simulation Constraints and Choice of Hyper-parameters}

Tables \ref{tab:hyper_Poisson} and \ref{tab:hyper_heat} summarize the hyper-parameters used in our simulations. The major computational overhead during each training epoch arises from the allocation of Random Access Memory (RAM) to store all operations performed during the simulation. These stored operations are subsequently used to compute the loss function and optimize the network weights. After each epoch, the RAM is cleared to manage memory efficiently.

Given this constraint, a careful balance must be maintained between the number of collocation points, the depth of the neural network, and the cutoff dimension. In both simulations, we employed a 4-layer quantum neural network with 2 qumodes. The first two layers allow interaction between the qumodes, while in the final two layers, each qumode acts only on its own state, represented in phase space by the position (\(\hat{X}\)) and momentum (\(\hat{P}\)) quadratures. This architecture was determined through trial and error during preliminary simulations.

Regarding the cutoff dimension, a value of 15 was sufficient to achieve accurate simulations for both equations. 

The number of collocation points (\(n\)) per training epoch is a crucial hyper-parameter that significantly impacts simulation accuracy. To maximize performance, \(n\) was chosen to fully utilize the available 64 GB of RAM. Additionally, \(n\) was set to \(2^k + 2\), ensuring optimal scrambling of \(2^k\) points by the Sobol algorithm, while two additional points were fixed at the boundary conditions.

\begin{table*}[ht]
\renewcommand{\arraystretch}{1.3}    
\setlength{\tabcolsep}{12pt}          
\centering
\begin{tabular}{lll}
\hline
\textbf{Hyper-parameter}           & \textbf{Value} & \textbf{Description}                                        \\ \hline
Multi-qumode layers            & 4              & Number of QNN layers acting on multiple qumodes                \\ \hline  
Single-qumode layers           & 4              & Number of QNN layers acting on a single qumode                 \\ \hline
Cutoff dimension               & 10             & Dimension of the truncated Hilbert space                       \\ \hline
Learning rate                  & 0.01            & Initial learning rate for the optimizer    
\\ \hline
Optimizer type                 & Adam           & Optimization algorithm used for training                        \\ \hline
Epochs                         & 5000           & Number of training iterations                                      \\ \hline
Collocation points     & 258             &  Number of points for training                        \\ \hline
\end{tabular}
\caption{Hyper-parameters used for the 1D poisson equation simulation.}
\label{tab:hyper_Poisson}
\end{table*}

\begin{table*}[ht]
\renewcommand{\arraystretch}{1.3}    
\setlength{\tabcolsep}{12pt}          
\centering
\begin{tabular}{lll}
\hline
\textbf{Hyper-parameter}               & \textbf{Value} & \textbf{Description}                                           \\ \hline
Multi-qumode layers                   & 2              & Number of QNN layers acting on multiple qumodes                \\ \hline  
Single-qumode layers                  & 2              & Number of QNN layers acting on a single qumode                 \\ \hline
Cutoff dimension                      & 20             & Dimension of the truncated Hilbert space                       \\ \hline
Learning rate                         & 0.01           & Step size for the optimization process                         \\ \hline
Optimizer type                        & Adam           & Optimization algorithm used for training                       \\ \hline
Epochs                                & 1000            & Number of training iterations                                  \\ \hline
Spatial collocation points            & 18             & Number of collocation points along the spatial dimension       \\ \hline
Temporal collocation points           & 10             & Number of collocation points along the temporal dimension      \\ \hline
\end{tabular}
\caption{Hyper-parameters used for Heat equation simulation.}
\label{tab:hyper_heat}
\end{table*}

\section{\label{app:optimizers}A benchmark on different classical optimizers}

\begin{figure}
    \centering
    \includegraphics[width=\linewidth]{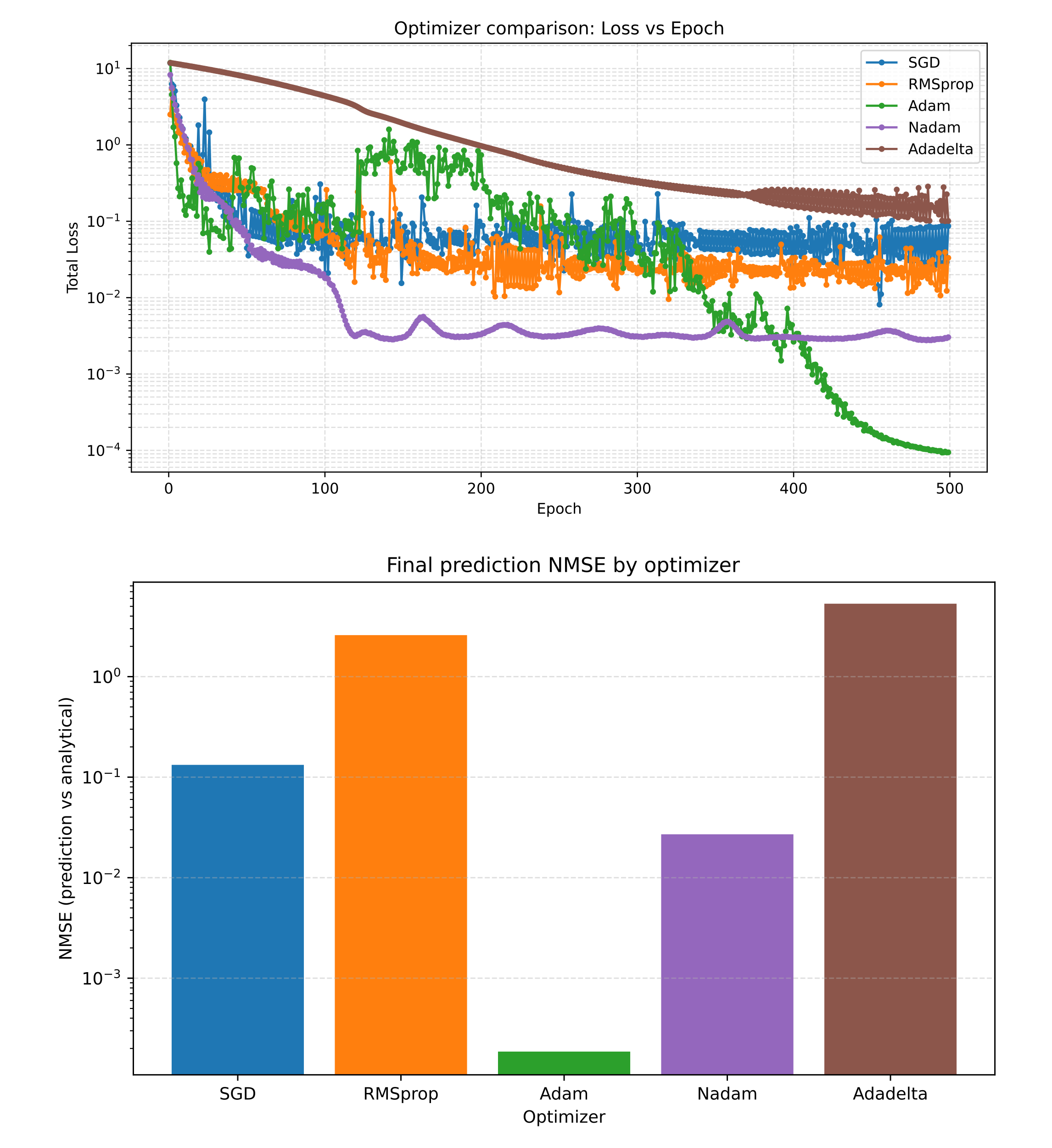}
    \caption{Benchmarking different the QPINN to solve the Poisson equation using different optimizers. In the top figure the loss function over epochs using different optimizer for 500 epochs. In the bottom figure we can see the value of the final error metric in the different cases. We can see that the Adam optimizer is the fastest and most precise. }
    \label{fig:optimizer_benchmarks}
\end{figure}

In this Appendix, we showcase how the outcomes from the simulations vary adopting different optimizers. More specifically, focusing on the 1D Poisson equation, we show how the loss function evolves by the number of epochs  and ì the final error of the solution trained with the different optimizers (Figure \ref{fig:optimizer_benchmarks}). We show that the Adam optimizer is the one with the fastest convergence towards the desired solution.

\newpage
\newpage


\end{document}